\documentclass[aps,pre,reprint,superscriptaddress,longbibliography]{revtex4-1}

\usepackage{graphicx}
\usepackage{times}
\usepackage{epstopdf}
\usepackage[english]{babel}
\usepackage{amssymb}	
\usepackage{amsfonts}
\usepackage{amsmath}
\usepackage{mathrsfs}
\usepackage{mathtools}
\usepackage{times}
\usepackage{adjustbox}
\usepackage{algorithm}
\usepackage[noend]{algpseudocode}
\usepackage[mathscr]{euscript}

\begin{document}



\title{Inference and Influence of Large-Scale Social Networks Using Snapshot Population Behaviour without Network Data
}


\author{Antonia Godoy-Lorite}
\affiliation{EPSRC Centre for the Mathematics of Precision Healthcare, Department of Mathematics, Imperial College London, London SW7 2AZ, United Kingdom}
\affiliation{Centre for Advanced Spatial Analysis. University College London, First floor, 90 Tottenham Court Road, London, United Kingdom} 

\author{Nick S. Jones}
\affiliation{EPSRC Centre for the Mathematics of Precision Healthcare, Department of Mathematics, Imperial College London, London SW7 2AZ, United Kingdom}
\affiliation{Department of Mathematics, Imperial College London, London SW7 2AZ, UK}


\date{\today}
\begin{abstract}

Population behaviours, such as voting and vaccination, depend on social networks. Social networks can differ depending on behaviour type and are typically hidden. However, we do often have large-scale behavioural data, albeit only snapshots taken at one timepoint. We present a method that jointly infers large-scale network structure and a networked model of human behaviour using only snapshot population behavioural data. This exploits the simplicity of a few parameter, geometric socio-demographic network model and a spin based model of behaviour. We illustrate, for the EU Referendum and two London Mayoral elections, how the model offers both prediction and the interpretation of our homophilic inclinations. Beyond offering the extraction of behaviour specific network structure from large-scale behavioural datasets, our approach yields a crude calculus linking inequalities and social preferences to behavioural outcomes. We give examples of potential network sensitive policies: how changes to income inequality, a social temperature and homophilic preferences might have reduced polarisation in a recent election.

\end{abstract}


\maketitle


\section{Introduction}


\begin{figure*}[t!]
\hspace*{-0.7cm}\includegraphics[width=2\columnwidth]{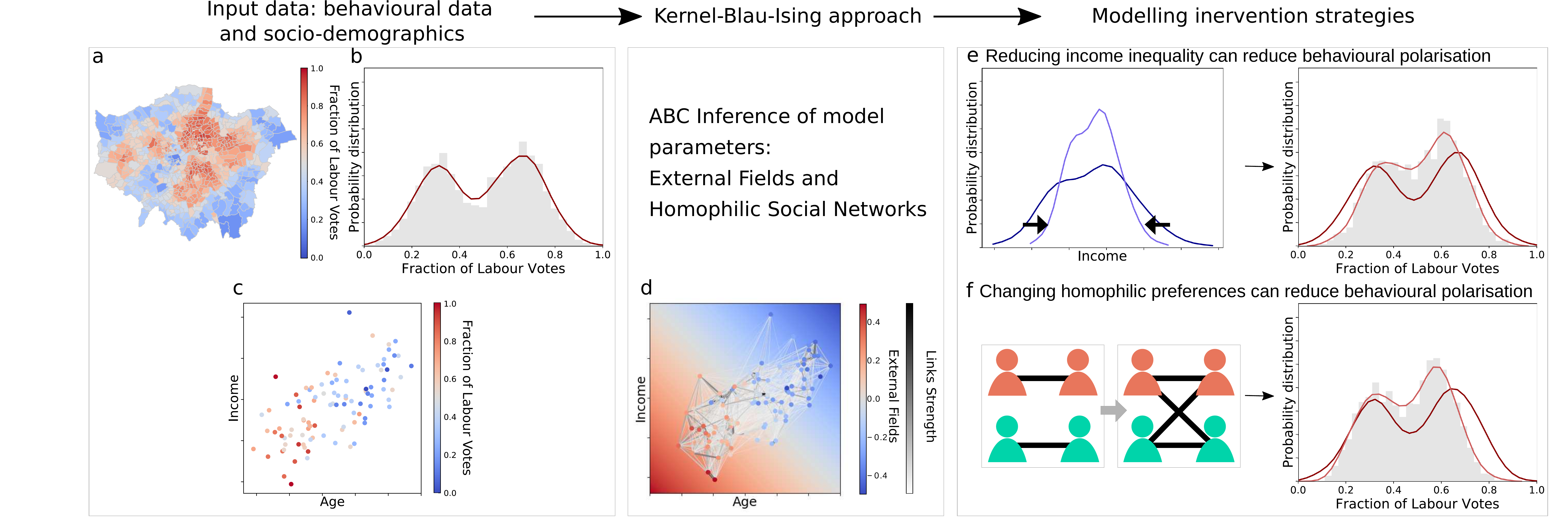}
\caption{\textbf{Outline of the kernel-Blau-Ising methodology.} Input data consist of: aggregated behavioural data at for different geographical areas and socio-demographic variables (age, income, education, etc.) associated to those areas (from census data). \textbf{(a)} Heatmap of (hypothetical) behavioural data in Greater London, in this case electoral outcomes, where red represents 100\% votes to Labour and blue represents 100\% votes to Conservatives. \textbf{(b)} Probability distribution of behavioural outcomes in (a). \textbf{(c)} Blau space representation of the behavioural outcomes spanned by socio-demographic characteristics (e.g. age, income). \textbf{(d)} Blau space representation of Kernel-Blau-Ising approach using input data in (1), and learning parameters: the External Fields, which account for the general trends, e.g. older people are more likely to vote Conservatives than younger people; and the Social Network that connects population according to their distances in the Blau space and their homophilic preferences. Once the model parameters are learnt, we can further estimate how changes and interventions affect behavioural outcomes. Examples of potential network-sensitive intervention strategies: how changes to income distribution \textbf{(e)} and homophilic preferences \textbf{(f)} can reduce behavioural polarisation.}
\label{fig:graphicalAbs}
\end{figure*}

Human behaviour, from voting preferences to vaccine sentiments, can depend on social networks \cite{zhang19}. While we have huge, high-quality, social-scientific datasets linking the behaviour of individuals to their individual circumstances (from censuses through health surveys to voting outcomes) it is extremely costly, or even impossible, to have direct access to the social networks on which this behaviour is articulated. The need to understand social network structure and how it shapes behaviour appears acute: there are concerns about both the role of social networks in health from vaccine refusal to obesity \cite{onnela10,larson16,christakis07}, and the recurring notion that our societies are becoming excessively polarised \cite{johnston04}. By accessing social-networks and the behavioural dynamics they support, we could also improve our perturbative understanding: clarifying how changes in social inequalities might change health and social polarisation. 

Given the need for social-network data, it follows that there has been immense scientific excitement about data from large networking platforms from Twitter to Mobile Phones \cite{lazer09,kosinski15}. It is, however, widely acknowledged that technology-platform data has numerous practical issues. A leading concern is whether technology-dependent network datasets give a true indication of the social networks on which society-relevant behaviours, like smoking or voting, depend; it is likely, instead, that different behaviours are spread on different aspects of our social networks \cite{cowan17}.  Technology-dependent network datasets are commercially sensitive and so are hard to access and share, are often available for limited time-spans or spatial extents, and, indeed, specific platforms themselves are unlikely to exist indefinitely: this creates concerns for reproducibility and generalizability. The most substantial issue, however, which must limit all such efforts in the future, is the immense privacy implication of large-scale social-network data: social network data is hard to anonymize \cite{narayanan09,montjoye18}. 
An alternative route to using data from technology platforms is to use conventional surveys-surveys though, beyond issues with scalability, it is often a challenge to identify whether the inferred networks are the network on which a particular behaviour is articulated \cite{mccarty19, mcpherson19,hoffmannthesis}. A third established route is to attempt to infer network models through e.g.  
%
%
time-series data \cite{timme14,peixoto19,schaub19,hoffmann20}. These approaches typically assume repeated observations of individual-level data; unfortunately human behaviour, such as voting or smoking, is often sampled at a single point in time. 

While it might thus seem challenging to access behaviour-specific social network structure there is one distinctive feature of social data which assists network inference: unlike many networked systems, censuses provide socially-relevant coordinate information for individual nodes. Peter Blau postulated an intuitive and powerful theory for social structure such that each individual in a society can be considered as being a point in a high-dimensional space (with dimensions like age, gender and income) where the rates of connection between individuals are driven by homophily and depend on their relative separation in the space \cite{Blau77,McPherson01,hipp09}. This homophily suggests that we can consider individuals to form social network links conditional on their separation in social space: typically modelled by a soft random geometric graph \cite{hoffmannthesis,hipp09,hoff02}. Beyond information about the coordinates of individuals, large-scale health and voting datasets give us snap-shot information about the behaviour of nodes. There is well developed theory to capture discrete choices \cite{mcfadden01} which in turn has links to finite-temperature linear threshold models of influence \cite{lynn16,kempe15} and Ising models \cite{brock01,gallo09,galam91}. This paper exploits a merger between Blau's geometric view of social structure and Ising models of behaviour to infer kernels for soft-geometric random graph models of social networks; we name this a kernel-Blau-Ising (KBI) model.

Our model allows behaviour to depend on both social circumstances and the behaviour of neighbours in a social-network. We illustrate this conceptualisation in Fig. \ref{fig:graphicalAbs}. It takes as input a set of individuals in a social-space and invokes both a Ising model for behaviour and a simple (soft  random geometric graph) model for how distances in a social space affect the chance of connections (friends, confidants, etc.). The simplicity of our model means we can use it to infer network parameter values for simulated data which carries no network information but only a snapshot of system behavioural state and node coordinate information. We illustrate our results for the EU Referendum and two London Mayoral elections, using only census data and voting outcomes, where we infer network parameter values consistent with the literature and we are able to predict missing/suppressed voting data. Finally our model allows us to quantify, model-dependently, the potentially depolarising effects of shifts in social connectivity preferences (e.g. eliminating income or age homophily) and social coordinates (e.g. reducing income inequality).

\section{An interpretable generative model for both population behaviour and social networks in Blau space}
\begin{figure}[t!]
\includegraphics[width=\columnwidth]{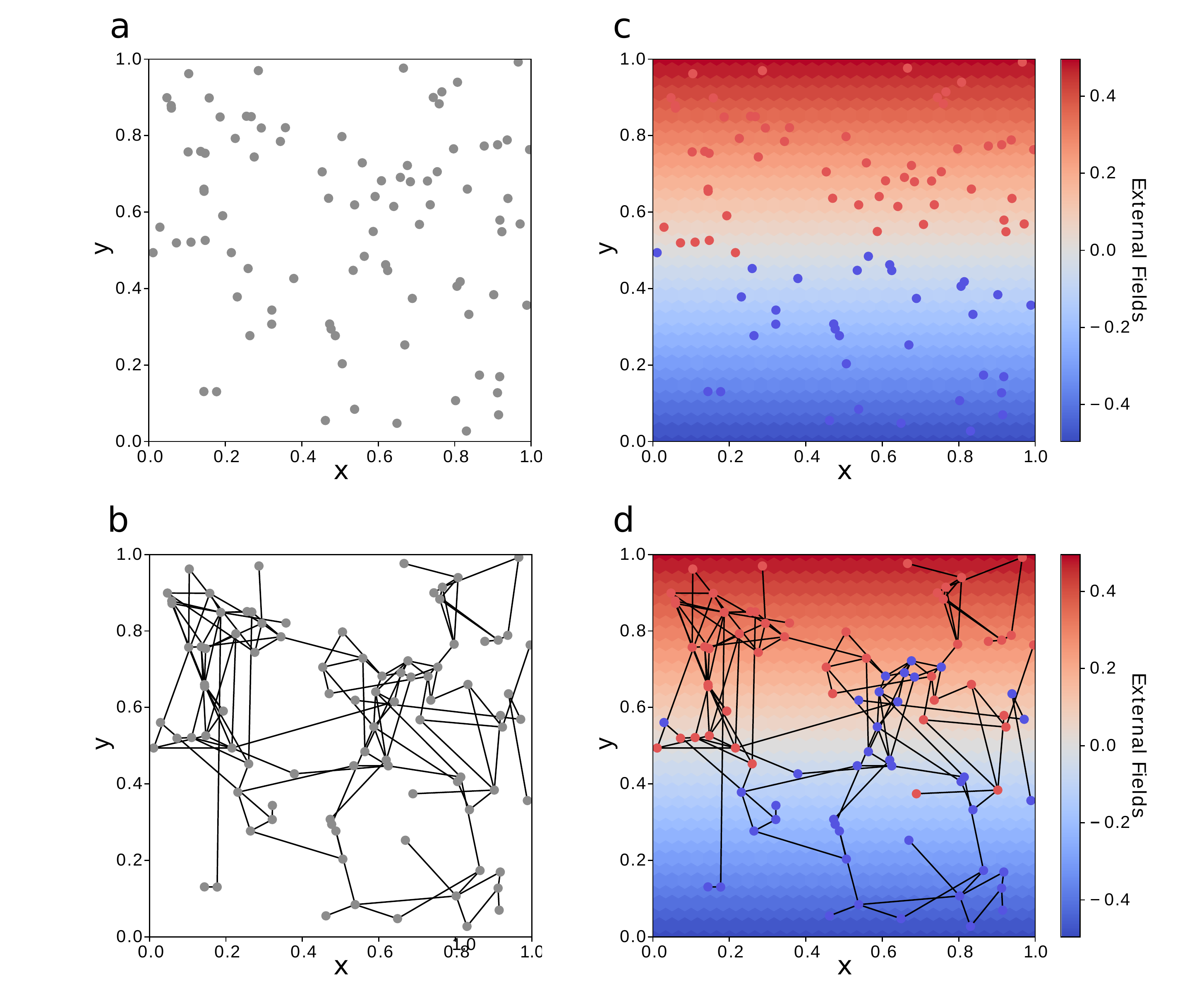}
\caption{\textbf{Generative process for spin configurations balances social/spatial fields and social network effects}. Panel \textbf{(a)} shows the coordinates of nodes in a two dimensional Blau space (e.g. $x,y$ is age vs income). In \textbf{(b)} we show a realisation of a Soft Random Geometric Graph (SRGG) from the connectivity kernel parameters $\theta_x=\theta_y=10$ ($\theta_0=0$). Panel \textbf{(c)} shows a spin configuration for a linear External Field (EFs) in the y-axis and low thermal noise ($\beta=100$), where the spins are aligned with the local EFs. Finally, in \textbf{(d)} we show the spin configuration under the same External Fields as in (c), also with low thermal noise ($\beta=100$), but now the spins are connected according to the SRGG in (b). We see how spins that in (c) aligned with the EFs have now changed their orientation to align more with their neighbours.}
\label{fig:cartoon}

\end{figure}

We deploy a generative model for population behaviour where the behaviour of individuals is partly determined by their social coordinates (as would be standard in logistic regression from survey data, e.g. regressing vaccine refusal on age and income); and partly determined by the behaviour of their neighbours on a social network as would be standard in socio-physics models \cite{gallo09,fedele13}. Regarding the social network, we also deploy a generative model for social networks where the chance that individuals have a social connection depends on their proximity in social space. In our model we consider binary social outcomes (for example voting Conservative/Labour or being smoker/non-smoker), but it is possible to extend the model to a discrete set of possible outcomes by using a Potts instead of an Ising model (we will keep to binary outcomes for simplicity). We use an Ising-like model (or Binary Markov Random Field) to model population social outcomes, but instead of locating spins in a regular grid, in our approach spins will be embedded in a multidimensional Blau space (where dimensions are socio-demographic variables and geographical coordinates) and social links between individuals occur with a probability depending on their separation in the Blau space.

\emph{Social network model:} We have $N$ individuals each embedded in a K-dimensional Blau space, where vector $z_i \in \mathbb{R}^K$ encodes the ith individual's coordinates in the Blau space representing her age, income, residential coordinates, etc. (see Fig. \ref{fig:cartoon}A showing random coordinates of nodes in a 2D Blau space).  We connect individuals through a soft random geometric graph (SRGG) \cite{penrose16} according to a connectivity kernel function which depends on distances in the Blau space and the kernel parameters (see Fig. \ref{fig:cartoon}B for an example of a SRGG).  This model makes it easy to simulate realistic networks with clustering \cite{handcock07}, although it does not explicitly build in other real social networks properties such as heavy-tailed degree distributions. Nonetheless, it is a well-established model for generating social networks that provides interpretable results \cite{hoff02,handcock07,hoff08}. 

We coded connections between spins in an adjacency matrix where $A_{ij}=1$ if $i$ and $j$ are connected and $0$ otherwise, with $A_{ij}$ Bernoulli distributed with a connectivity kernel $\rho$, 
\begin{equation}
P(A_{ij}=1|z_i,z_j,\boldsymbol{\theta})=\rho(z_i,z_j,\boldsymbol{\theta}).
\label{eq:Aij}
\end{equation}
We choose our connectivity kernel to be a logistic sigmoid function as they have been successfully used for the inference of connectivity kernels on ego networks \cite{hoffmannthesis} and in latent-space inference \cite{hoff02,handcock07,hoff08}.
\begin{equation}
\rho(z_i,z_j,\boldsymbol{\theta})=\frac{1}{1+\exp(d_{ij})}; \:
d_{ij}=\theta_0+\sum_{k=1}^{K} \theta_k |z_{ik}-z_{jk}|,
\label{eq:kernel}
\end{equation}
where $d_{ij}$ is interpreted as the distance in the Blau space, $\theta_0$ is a bias term that accounts for the overall connectivity density regardless of the distances in the Blau space, and $\theta_k$ is the connectivity coefficient of Blau dimension $k$ which weights the contribution of distances in the $k$ dimension to the overall distance. The connectivity coefficients ($\theta_k, \:k=\{1,..,K\}$) measure homophily in the Blau space, so that the larger $\theta_k$ becomes the stronger the homophily in that dimension (connections becoming more localised in that dimension).  Importantly, the constant bias term $\theta_0$ allows rescaling of system size, since it accounts for density changes without modifying values of the connectivity kernel parameters ---see Supplementary Material (SM) Section S2. The connectivity kernels induce an interpretable semi-metric for connections on the Blau space \cite{hoffmannthesis} and can be used to generate particular realisations of SRGGs.

\emph{Behavioural model:} Our Ising model to generate spin configurations is as follows. Each individual $i$ in the population has a spin associated $\sigma_i=\{-1,1\}$ encoding her binary social outcome, so that a spin configuration is $\sigma \in \{-1,1\}^{N}$. 
The spin orientation depends on the external fields (EFs) and the other spins they are connected to in the network. As is common for conventional social statistics (logistic regression with linear dependence on the covariates), we model the EFs as linear fields in each dimension of the Blau space, where the linear coefficient in each dimension of the Blau space $k$ is $h_k$, so that the individual spin interaction with the EFs is the scalar product $h\cdot z_i=\sum_k h_k z_{ik}$. The spins interact with the external fields depending only on their coordinates so that they tend to align with the EFs (see Fig. \ref{fig:cartoon}C). 

The energy of a spin configuration $\sigma$ is given by the Hamiltonian function,
\begin{equation}
H(\boldsymbol{\sigma},\boldsymbol{h},J,A(\boldsymbol{\theta}))=-\sum_{i} \left( \sum_k h_k z_{ik} \right) \sigma_i - J \sum_{ij}A_{ij} \sigma_i \sigma_j.
\label{eq:hamiltonian}
\end{equation}
where $h_k$ is the EFs linear coefficient in dimension $k$ of the Blau space, $J$ is the connection strength and $A_{ij}$ is the adjacency matrix. We can add an homogeneous field $h_0$ which is felt by the whole population regardless of their coordinates. However, in the cases we consider in the following it is reasonable to set $h_0$ to zero. 
The configuration probability is given by the Boltzmann distribution with inverse temperature $\beta = 1/T, \: \beta \geq 0$:
\begin{equation}
p(\boldsymbol{\sigma}|\beta,\boldsymbol{h},J,A(\boldsymbol{\theta}))=\frac{e^{-\beta H(\boldsymbol{\sigma},\boldsymbol{h},J,A(\boldsymbol{\theta}) )}}{ Z(\beta,\boldsymbol{h},J,A(\boldsymbol{\theta}))},
\label{eq:Boltzmann}
\end{equation}
and the normalisation constant,
\begin{equation}
Z(\beta,\boldsymbol{h},J,A(\boldsymbol{\theta}))=\sum _{\boldsymbol{\sigma'} \in \Omega }e^{-\beta H(\boldsymbol{\sigma'},\boldsymbol{h},J,A(\boldsymbol{\theta}))}
\label{eq:partitionfunction}
\end{equation}
is the partition function, where the sum is over all possible spin configurations, $\Omega$, which for and Ising model are $2^N$ terms. The configuration probabilities $p(\boldsymbol{\sigma}|\beta,\boldsymbol{h},J,A(\boldsymbol{\theta}))$ represent the probability that, in equilibrium, the system is in a state with configuration $\sigma$.  Fig. \ref{fig:cartoon}D gives an example of a spin-assignment which has been generated conditional on a particular network structure --in our case a SRGG from a particular connectivity kernel. Vitally, the spins of the nodes are not exclusively determined by either the external fields or network structure.




\section{Inference method for Model Parameters}
Given a record of a social outcome $\boldsymbol{\sigma} \in \{-1,1\}^{N}$ together with the population Blau space coordinates $z \in \mathbb{R}^{N \times K}$, our goal is to find the most plausible model parameters $\Theta=\{\beta,\boldsymbol{h},J,\boldsymbol{\theta\}}$.  
For the Ising model the partition function $Z$ cannot be computed even for small systems, since it requires the computation of $2^N$ terms. A further challenge is that $Z$ itself depends on the model parameters and would need to be recomputed for each possible parameter set. In this case the inference is called doubly-intractable \cite{murray06} and Markov chain Monte Carlo is challenging. 


As an alternative likelihood-free method (that nonetheless avoids mean-field approximation) we use approximate Bayesian computation (ABC), which has been applied to a wide spectrum of problems with intractable likelihoods \cite{marin12,beaumont02}. 
In Algorithm \ref{algm:ABC} we show our rejection-based ABC inference method for the model parameters. We suppose we have priors, $\pi(\Theta)$, on possible parameter values. Importantly, on \textit{lines 4,5} the generation of a spin configuration requires two steps: (i) the generation of a soft random geometric graph (SRGG) from the connectivity kernel Eq.\eqref{eq:kernel} given the node coordinates; (ii) the generation of the spin configuration conditional on the graph generated in (i) and for the parameters of the Hamiltonian in Eq.\eqref{eq:hamiltonian}.

\begin{algorithm}[H]
\caption{ABC rejection algorithm}
\begin{algorithmic}[1]

\For{$i=1$ to $N$}
    \Repeat
        \State Draw independent proposal $\Theta' \sim \pi(\Theta)$
        \State Generate $A' \sim \rho(\boldsymbol{z},\boldsymbol{\theta}')$
        \State Generate $\boldsymbol{\sigma'} \sim p(\cdot|A',\Theta')$ 
	\Until{$\left\| \eta(\boldsymbol{\sigma'},\boldsymbol{\sigma})  \right\| < \epsilon$}
		\State Set $\Theta_i \gets \Theta'$
\EndFor

\end{algorithmic}
\end{algorithm}
\label{algm:ABC}


We use Glauber dynamics to generate  spin configurations $\boldsymbol{\sigma'}$ for any combination of the model parameters $\Theta'$ from Boltzmann distribution. Importantly, in order to improve the efficiency of the ABC rejection algorithm, we define a set of lower-dimensional summary statistics. We summarise spins that share the same coordinates $z$ by the fraction of spins down (or up) $S_z=\frac{1}{n_z} \sum_{i|z_i \in z}\delta(\sigma_i,-1)$,  where $n_z$ is the number of individual spins at coordinate $z$ of Blau Space. If there are $C$ different Blau space coordinates populated with spins, the summary statistic would be such that $S(\boldsymbol{\sigma})\in [0,1]^C$, with ideally $C \ll N$. Therefore, We approximate our posteriors by $p(\boldsymbol{\sigma}| (\left\| \eta(S(\boldsymbol{\sigma'}),S(\boldsymbol{\sigma}))  \right\| < \epsilon ) )$, where $\eta(S(\boldsymbol{\sigma'}),S(\boldsymbol{\sigma}))$ measures the discrepancy between $\boldsymbol{\sigma'}$ and $\boldsymbol{\sigma}$ after they are summarised with function $S(\boldsymbol{\sigma'})$. We define the distance between $S(\boldsymbol{\sigma})$ (summary statistics of the original social outcomes data) and $S(\boldsymbol{\sigma'})$ (observational data of the spins configuration generated from $\Theta'$) as, 
\begin{equation}
   \eta(S(\boldsymbol{\sigma'}),S(\boldsymbol{\sigma}))= \frac{1}{N}\sum_{z=1}^C n_z |S_z-S'_z|,
   \label{eq:eta}
\end{equation} 
where $|S_z-S'_z|$ is the absolute difference between the fraction of spins down in the observed spin data and the generated spins, so that the distance is zero only if $S_z=S_z'\: \forall z$. The distance $\eta(S',S)$ can also be considered as a weighted mean absolute error (WMAE). Significantly, our observational data itself will be aggregated in the form of fractions of spin-up nodes in different small spatial regions (e.g. proportions of smokers or voters in different small spatial patches  -- as in Section \ref{sec:ResVoting}). Since we are precisely attempting to simulate our observational data, and given that all spins with the same coordinates are statistically indistinguishable (Eq.\ref{eq:hamiltonian}), vitally, $S(\boldsymbol{\sigma})$ are sufficient statistics and the ABC posteriors tend to exact Bayesian posteriors in the limit of $\epsilon \rightarrow 0$. For large-scale data applications (as in our voting illustration), population size can be rescaled through rescaling the connectivity kernel bias term $\theta_0$ in Eq.\ref{eq:kernel} (see SM Section S2). Since the ABC approximate posteriors of the model parameters are obtained through independent samples, it is straightforward to parallelise the algorithm so that the samples are obtained concurrently. For a more efficient sampling procedure sequential sampling schemes for ABC can also be used \cite{dutta17}. 

\section{Results}

\subsection{ABC Inference can estimate model parameters from synthetic data}
\label{sec:ResSynthetic}
\begin{figure}[t!]
\includegraphics[width=\columnwidth]{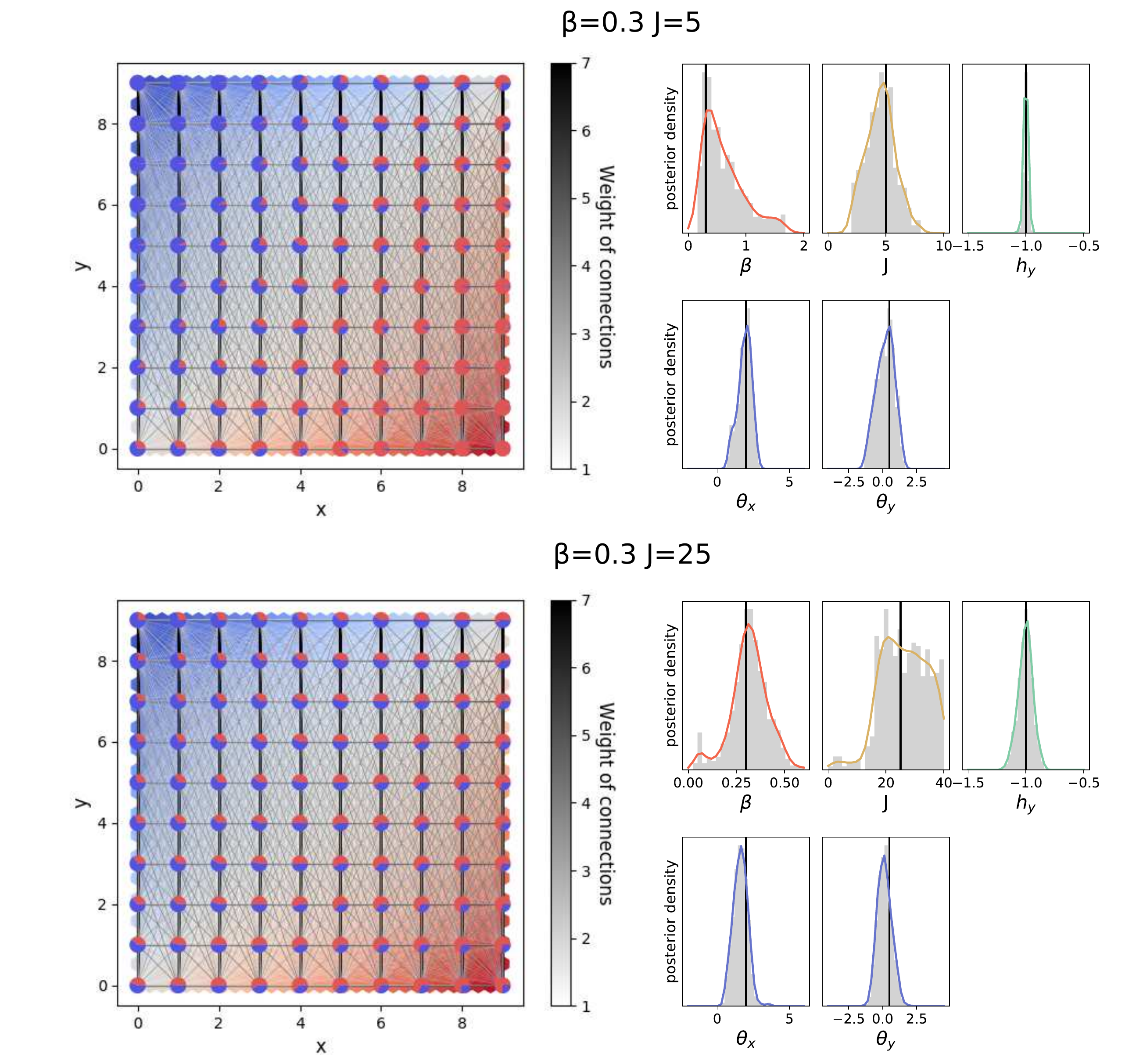}
\caption{\textbf{Inference allows the recovery of model parameters for synthetic snapshot data}. On the top left, synthetic data with a weak connectivity strength spin configuration (J=5), and on the bottom right a strong connectivity strength spin configuration (J=25); both for $\beta=0.3$ (see Fig. S3) and kernel parameters $\theta_0=9$, $\theta_x=2$ and $\theta_y=0.5$. There are a total of $N=10,000$ spins, with $100$ spins on each of the discrete coordinates on the grid where $x,y=\{0,1,..,9\}$ (for visualisation purposes links are aggregated at the coordinate level). To avoid coupling between certain model parameters we choose to set $h_x=1$ and $\theta_0=9$ for the ABC inference (see main text). We use uniform priors for $\beta \in [0,2]$, $h_y \in [-1.5,0.5]$, $\theta_x \in [-0.5,4.5]$, $\theta_y \in [-2.5,2.5]$, and $J\in [0,10]$ in the $J=5$ scenario and $J\in [0,40]$ in the $J=25$ scenario. We show the ABC marginal posterior distributions for $500$ samples with the lowest distance $\eta(S',S)$ in Eq.\ref{eq:eta}. The samples are visualised using histograms in grey and Gaussian kernel density estimates as solid lines, and the vertical lines correspond to the real values used to generate the synthetic spin configuration. On the right, for $J=25$, it shows that the ABC inference is not able to distinguish between configurations above a given values of $J>J_{\mathrm{aligned}} $ since all spins that are connected are already aligned. The ABC inference algorithm accurately estimate the connectivity kernel parameters without using network data for synthetic systems.
}
\label{fig:synthetic}
\end{figure}
We tested the ability of our ABC inference method to recover the known parameters, of both the network and behavioural processes, for synthetic snapshot data. We note that, in a manner distinct from other inverse-Ising or network inference approaches \cite{fedele13b,fedele17}, 
we are not seeking to recover unique network links, do not observe each node state (only coarsened observations) and we will not use time-series data. As it is reasonable for social data, we suppose that we are given access to information about the social coordinates of nodes/individuals. Our simple, but justified, model structure allows us to extract information from very limited datasets composed of snapshot behavioural data and census information. Given that survey and census demographic-variables are typically ordinal or categorical, we use ordinal data in our experiments with synthetic data. We performed the ABC rejection method for two different combinations of connection strengths at a temperature below the corresponding critical temperature $\beta> \beta_c$ (see Fig. S3). We found that for a given temperature $\beta$ there is a value of $J_{\mathrm{aligned}}$ where every pair of nodes that are connected are aligned ($J_{\mathrm{aligned}}$ changes for the different $\beta$ values). Therefore, any $J>J_{\mathrm{aligned}}$ has the same distribution over spin configurations where connected spins are aligned. We test our inference method in two different synthetic data scenarios, one with strong connection strength $J_s>J_{\mathrm{aligned}}$, and another with weak connection strength $J_{w}<J_{\mathrm{aligned}}$.

In Fig. \ref{fig:synthetic} we show the ABC posteriors for the different scenarios, with $J=5$ (weak) and $J=25$ (strong). Importantly, we have set two model parameters in the inference: the $x$ axis external field $h_x$ ($h_x=1$) and the connectivity bias term $\theta_0$. Regarding $h_x$, from  Eq.\ref{eq:Boltzmann} we see that the inverse temperature parameter $\beta$ multiplies the linear external fields $\boldsymbol{h}$, thus there is one degree of freedom that we choose to reduce by setting $h_x=1$ without violating any constraint. Regarding the connectivity bias term $\theta_0$  in Eq.\ref{eq:kernel}, there is a coupling between the connection strength $J$ and the bias term $\theta_0$ in the Hamiltonian Eq. \ref{eq:hamiltonian}. This implies that, in some regimes, the same spin configurations can be generated by different combinations of these two factors, i.e, a spin configuration could be a result of a certain combination of low connectivity density (large $\theta_0$) and strong connection strength $J$ and vice-versa, high connectivity density (small $\theta_0$) and weak connection strength ----this trade-off does not hold when the link density is very low, with almost no links, or when it is very high, such that it become almost a complete network, see SM Fig S2. Therefore, we set the connectivity bias term $\theta_0=9$ in the inference (such that the average degree is $\kappa \sim 2$, a parameter choice discussed in the next section). 
For the ABC inference in Fig. \ref{fig:synthetic}, we show the ABC marginal posteriors for the best $500$ samples according to the datasets distance $\eta(S',S)$ Eq.\ref{eq:eta}. 
The distances corresponding to the spins configuration in Fig. \ref{fig:synthetic} are $\epsilon=0.036$, which can be also be seen as a weighed mean absolute percent error (WMAPE) lower than $3.6\%$.
Results show that all inferred parameters are consistent with the original values both for weak and strong connection scenarios. As expected, the posterior of the connection strength for $J=25$ accepts all values for $J>J_{\mathrm{aligned}}$. Considering the connectivity kernel parameters, we show we can recover posteriors consistent with $\theta_x$ and $\theta_y$ from one single observation of a spin configuration without using network data.

\subsection{Inferred parameters for Mayoral Elections and EU Referendum are consonant with homophilic tendencies and voting preferences.}
\label{sec:ResVoting}

\begin{figure*}
\includegraphics[width=2\columnwidth]{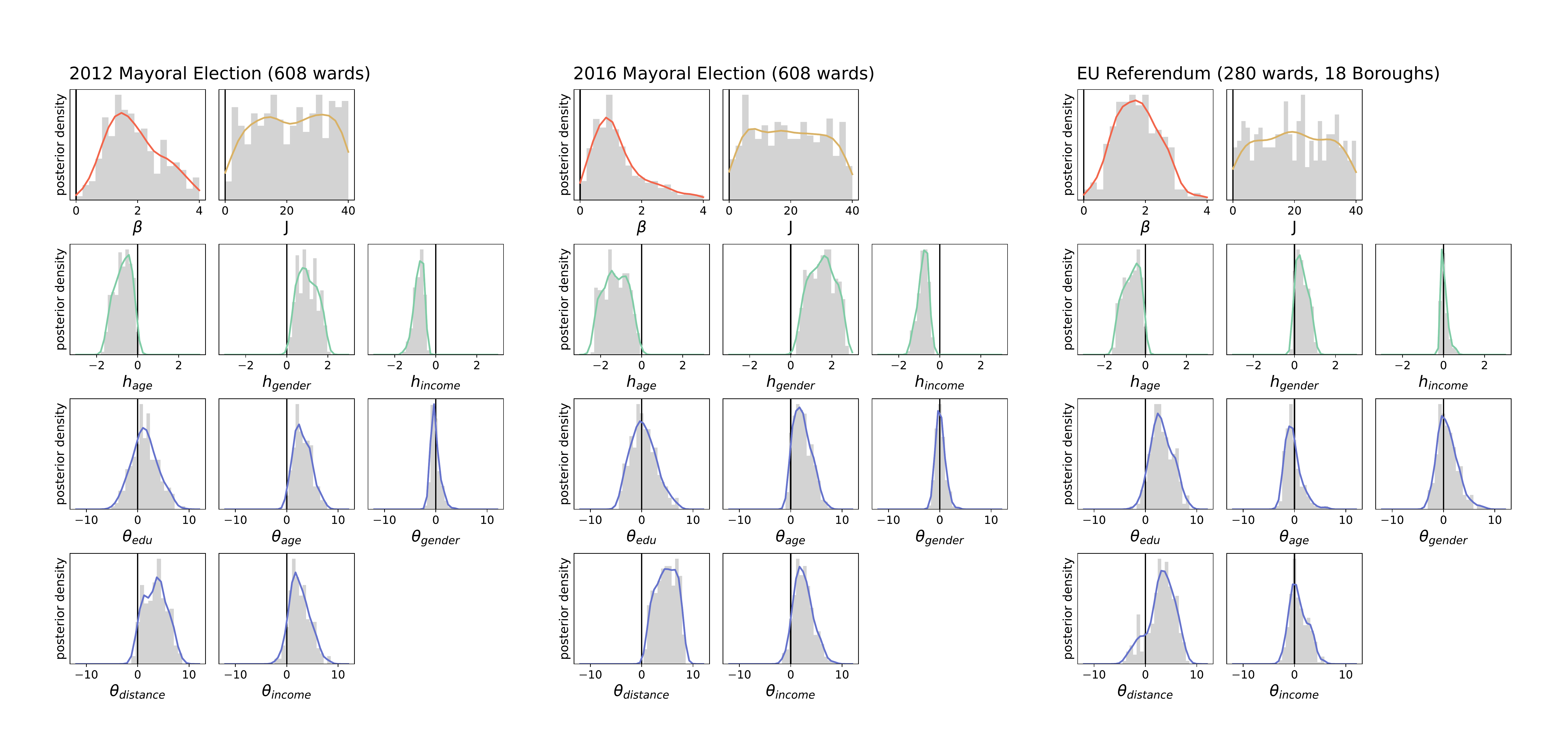}
\caption{
 \textbf{ABC marginal posteriors for London Mayoral Elections 2012 and 2016 and EU (Brexit) Referendum are consistent with known homophilic and political preferences}. We show ABC marginal posteriors for London Mayoral Elections 2012 and 2016 for $608$ electoral wards (See ABC marginal posteriors for Mayoral Elections 2012 and 2016 for $280$ electoral wards and 18 Boroughs Fig. S7). We show estimates for parameters: inverse temperature $\beta$ in red, the connectivity strength $J$ in yellow, the external field for age, gender and income, $h_{i};\: i=\{age, gender,income\}$, and the connectivity kernel parameters $\theta_i; \: i=\{education, age, gender, distance,income\}$. $h_{education}$ and $\theta_0$ as fixed (see main text). We use uniform priors for $\beta \in [0,4]$, $J \in [0,4]$, $h_{age} \in [-3,0.15]$, $h_{gender} \in [-0.3,3]$, $h_{income} \in [-1.6,0.7]$, $\theta_{education} \in [-7,12] $, $\theta_{age} \in [-5,12]$, $\theta_{gender} \in [-6,12] $, $\theta_{distance} \in [-7,11] $ and $\theta_{income} \in [-5,12] $. The ABC marginal posterior distributions are shown for $500$ samples with the lowest distance $\eta(S',S)$ (Eq.\ref{eq:eta}). We show the histograms of the ABC marginal posteriors in grey, as a solid line the Gaussian kernel density estimates, and the vertical lines correspond to $0$ value. The parameter estimates show that the connectivity kernel parameters of the two Mayoral Elections are similar but considerably different from EU Referendum connectivity kernel. Unlike the Mayoral Elections, the EU referendum does not show homophilic signal for age, gender and income but only for distance and education, even though the EU referendum took place a month after Mayoral Election 2016.
\label{fig:Bremay280}
}
\end{figure*}


\begin{figure*}
\includegraphics[width=2\columnwidth]{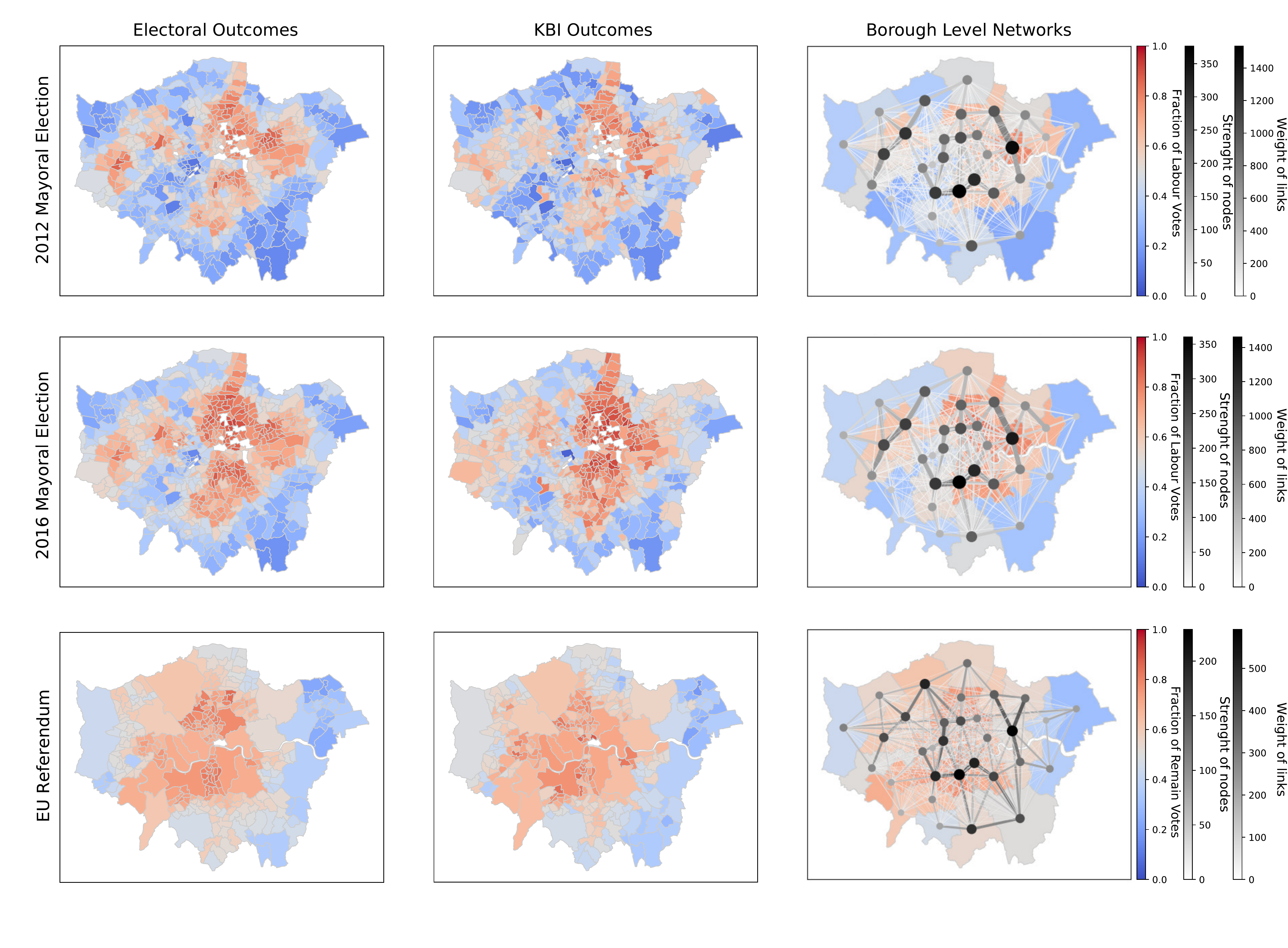}
\caption{
 \textbf{KBI accurately reproduces electoral outcomes while inferred connectivity networks highlight differences between Mayoral Elections and EU referendum}. 
 The \textit{left} column show the real electoral outcomes and next to it the \textit{middle} column show inferred outcomes generated from the ABC posteriors with lowest data distance, for the three election in Greater London. For London Mayoral elections, the colour range goes from red for $S_w=1$ ward's fraction of Labour vote, and blue for $S_w=0$ or all ward's votes for Conservatives. For EU Referendum, in order to keep colour consistency, we represent in red $S_{w/B}=1$ fraction of ward/Borough votes for remain, and blue  $S_{w/B}=0$ for wards/Boroughs voting for leave. On the \textit{right} column, we show the connectivity networks from Fig. \ref{fig:Bremay280} ABC marginal posteriors of the kernel parameters aggregated at the Borough level for visualisation purposes (ward level social networks with $184,528$ links are hard to visualise). The colour map in the background is the real electoral outcomes at the Borough level. For the connectivity kernel we coloured in a linear range (white to black) the strength of the nodes and the weight of the links, also the size of nodes and links is linear with their strength and weight respectively. We have not visualised intra-Borough links but we have take them into account for the nodes strengths. We see that MEs connectivity networks are similar among them but different from the EU Referendum connectivity network. All the networks show pronounced spatial homophily leading to strong links between neighbouring Boroughs.
\label{fig:connectivity}
}
\end{figure*}

To demonstrate our  model and inference approach beyond synthetic data, we apply it to three electoral datasets in Greater London: 2012 London Mayoral Election, 2016 London Mayoral Election and EU (Brexit) Referendum in 2016. Our kernel-Blau-Ising model allows us to compare the different electoral outcomes using readily interpretable parameters, to visualise and compare the social connectivity structures and to further estimate interventions. We stress that our objective is illustrative and note, of course, that much more refined social models, e.g. allowing connections between the external fields, could be deployed. 

The electoral results are given as aggregated outcomes, where we know the total outcome for a specific area but we do not know the votes of individuals. The smallest areas in spatial resolution that we can get for electoral outcomes are the electoral wards --  $630$ electoral wards in Greater London from which, due to data mismatches, we can only use $608$ wards outcomes; for the EU referendum ward level data is missing for 18 Boroughs, so we use a combination of $280$ ward level outcomes and $18$ Borough electoral outcomes data (see SM Section S1). While census microdata is available at the ward-level, we define the Blau space coordinates of each electoral ward as the average value from census data in each Blau space dimension, which are: education, age, gender, wards centroid spatial coordinates and income (see SM Section S1 for details). This coarsening overestimates proximity within wards (and could conceal heterogeneities) --favouring more homogeneous behaviour-- however we note that although we lose information about individuals' connectivity by using aggregated data instead of individual's micro-data coordinates, we see that average coordinates at the ward level are heterogeneous (in SM Fig. S4 we show evidence of heterogeneous distribution of average wards' coordinates by bootstrapping census micro-data for each ward). We finally note that census data is from 2011, and London is a fast changing city, nonetheless we consider this 1-5 year gap adequate for our illustrative aim.


We now define distances in the Blau space and how we can return reasonable posteriors for real data. We define distance between two wards in the Blau space (Eq. \ref{eq:kernel}) as the absolute difference of their coordinate values for education, age, gender and income dimensions, and as the distance between centroid coordinates for the spatial distance (see SM Section S1). Before passing the distances to the inference algorithm, we standardise them by subtracting the mean distance and then dividing by twice their standard deviation for each Blau dimension \cite{gelman08}. The standardisation allows us to compare homophily kernel parameters among them and makes interpretation easier. Finally, for the ABC rejection algorithm we simulate a representative sample of spins in each electoral ward instead of simulating the whole population. Specifically, we keep the relative size of the wards population according to the census data (from a population of $N\sim 8,800,000$ individuals in Greater London we rescale the system to $N=60,683$ with a average of $100$ spins per ward). The population rescaling inevitably affects the bias term $\theta_0$ in Eq.\ref{eq:kernel} but for sufficiently large samples can provide an adequate approximation for the homophily kernel parameters (see SM Section S2). Given the coupling between the connection strength $J$ and the connectivity bias term $\theta_0$, we set $\theta_0=14$ -- which corresponds to a social network with an average degree of $\kappa \approx 2$, comparable with real data on the estimated number of ego-confidants \cite{McPherson06}, see also Fig. S2 where we show evidence of insensitivity of dimension-specific kernel parameters to changes in the degree of the network. As per the synthetic data, we fixed one of the external fields $h_{edu}$ to remove the degree of freedom in the Hamiltonian (Eq. \ref{eq:Boltzmann}) between the inverse temperature and the External Fields. We choose to set $h_{edu}=0.45$ based on the results for a simple multilinear regression (see SM Section S5), where we find that the education linear coefficient has the least variability when comparing values among the three election. In Fig. \ref{fig:Bremay280} we show the ABC marginal posteriors for $500$ samples with the lowest data distances $\eta(S'S)$, with distances (or WMAE) $\eta(S',S)<0.088$ for 2012 and 2016 Mayoral Elections and $\eta(S',S)<0.050$ for the EU Referendum. The EU Referendum ward outcome distribution is less polarised than the MEs outcome distribution (see SM Section S8), and less polarised outcomes are better captured by linear External Fields (see SM Section S6 for the inference only with EFs), hence, overall KBI performance is better for the EU referendum than for MEs.

Figure \ref{fig:Bremay280} shows that the marginal posteriors for the two London Mayoral Elections are qualitatively similar (except possibly the level of noise, indicated by $\beta$) with the Conservative party winning the 2012 MEs while the Labour party won the 2016 MEs (see Fig. S8 for the detailed distribution of outcomes for the three elections). 
However, the EU Referendum ABC marginal posteriors present some differences. In terms of the parameters of the model, the temperatures are sub-critical for the the three elections, meaning that spins are in the ordered phase (see in SM Fig. S3). Regarding the connection strength $J$, for the three elections the value of $J$ is large enough that all spins connected are aligned and the inference can not distinguish $J$ values larger than $J_{\mathrm{aligned}}$ as in the strong connection scenario in Fig. \ref{fig:synthetic} (our choice of $\theta_0$ is partly determining our inferred $J$). For the EFs, the three elections show the same sign for the linear coefficients in the different dimensions, with the difference that EU referendum EFs are closer to zero than for the two MEs. Notice that $h_{education}$ is set to $0.45$ so that more educated people tend to vote Labour/Remain (see SM Section S5). The observed EFs are in agreement with traditional two-party partisan voting socio-demographic tendencies in the UK: $h_{age}$ is negative, meaning that older voters prefer Conservative/leave, $h_{gender}$ is positive, meaning that men vote more Conservative/leave than women, and $h_{income}$ is negative meaning that the richer the more Conservative they vote; although for the EU referendum income EF is peaked around zero.

Regarding the connectivity kernel parameters, the resulting connectivity kernels are non-negative for the three electoral datasets, which is in agreement with the homophilous tendency of social relations \cite{McPherson01,smith14,wang13}. Importantly, the connectivity kernel parameters for the two MEs are similar, but differ from EU referendum kernel parameters. Regarding the similarities, distance homophily is  persistently the strongest homophily signal in the three electoral outcomes, consistent with other work \cite{hipp09,hoffmannthesis}. Apart from spatial distance homophily, MEs' kernels show positive signal (homophily) for age and income dimensions; while for the EU Referendum age and income ABC marginal posteriors are peaked around zero, but the education kernel parameter shows a positive signal that is not present in the MEs' kernels. Notably the 2016 Mayoral Election was held on the $5^{th}$ of May 2016, only 49 days before the EU Referendum on the $23^{rd}$ of June. Therefore, the changes in the social kernels we infer, if true, are not due to temporary evolution of the social connectivity structure, but rather indicate that different social ties were at play for the EU Referendum compared to Mayoral Elections.

In Fig. \ref{fig:connectivity} we show a comparison between the real and the KBI generated electoral outcomes from the ABC posteriors with lowest data distance, 
and the social networks (aggregated at the Borough level) for the three electoral datasets. The social networks are computed as the average weights for the networks generated from the connectivity kernel parameter in Fig. \ref{fig:Bremay280} at the ward level and then aggregated at the Borough level purely for visualisation purposes (there are $184,528$ social links at the ward level). The differences in the connectivity kernels between MEs and the EU referendum indicate that different aspects of the social network were at play in the Mayoral Elections vs at the EU Referendum. We also computed the median distance for all links (at ward level) in the electoral social networks. We find that the median distance between socially linked individuals is similar for the three elections, for 2012 ME this is $5.75$Km and for 2016 ME $4.90$Km, for EU Referendum $6.31$Km, where importantly, the median Borough diameter is $6.22$Km (median ward diameter is of $1.38$Km); this is consonant with spatial homophily observed by others \cite{hoffmannthesis,McPherson01}. Therefore, most of the social links we infer are inside the Borough and to first neighbours at the Borough level.



\subsection{The kernel-Blau-Ising model accurately predicts unobserved voting outcomes}
\label{Sec: predictions}
\begin{figure*}
\includegraphics[width=2.\columnwidth]{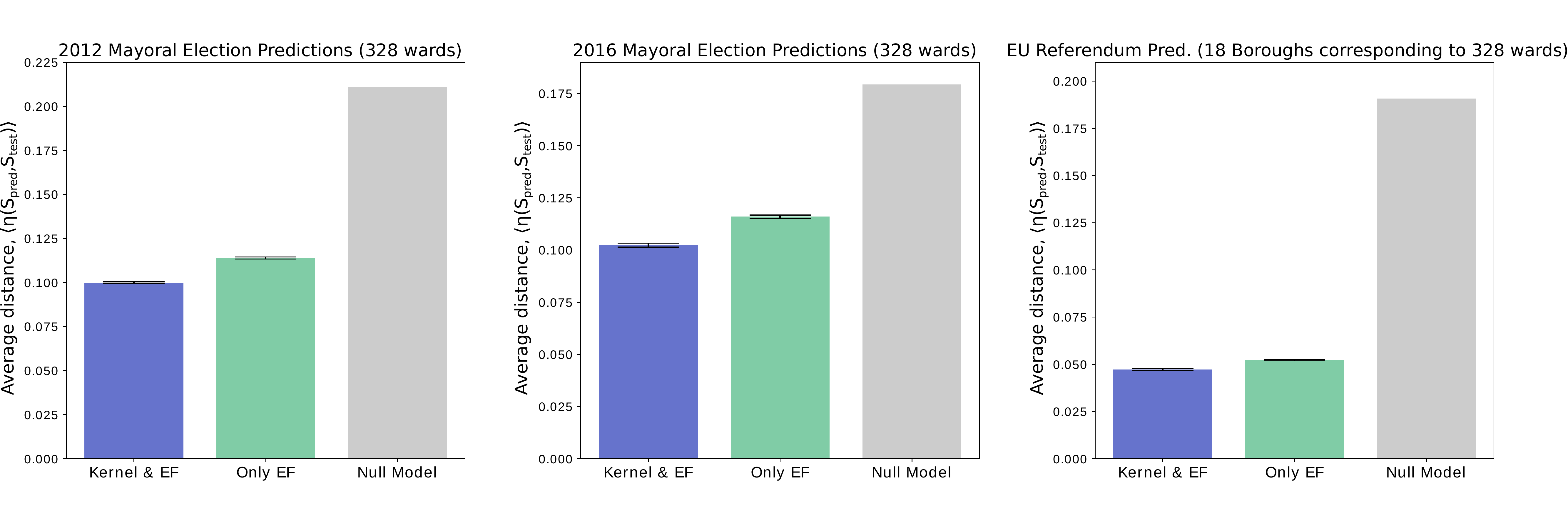}
\caption{\textbf{Kernel-Blau-Ising models which include the connectivity kernel outperform those using only External Fields in the three elections}. We measure prediction on 2012 and 2016 MEs and EU Referendum, where we train the models with $280$ observed wards' outcomes and make the predictions on $328$ test wards' outcomes (corresponding to $18$ Boroughs in EU Referendum), see SM Section S1. Each bar represents the average distance $\langle \eta(S_{pred}, S_{test}) \rangle$ between predicted wards'(boroughs') outcomes and test wards'(boroughs') outcomes over $100$ predictions from those parameters sets with the lowest distance in the training data. The error bars show the error of the mean. We show the predictions for the kernel-Blau-Ising model (KBI) in blue, for the model only using External Fields (without social kernel) in green, and we also show the predictions for a Null Model that predicts for all wards/Boroughs the weighted average computed from the training data in grey. KBI accurately predicts unobserved outcomes in the three elections, with the relative effect of social connections being lower in the EU Referendum.
\label{fig:predictability}
}
\end{figure*}

The distance measure between generated and real electoral outcomes $\eta(S',S)$ in Eq.\ref{eq:eta} for the three electoral datasets gives us a measure of the how accurately we can reproduce electoral outcomes. In the previous Section \ref{sec:ResVoting}, we showed that the KBI is able to accurately reproduce the electoral outcomes for the three elections, in particular with a weighted mean absolute percent error (WMAPE) below $8.8\%$ for MEs and bellow $5\%$ for the EU Referendum. In addition to these general results, we measure the predictive power of our approach. 

Motivated by the fact that we have missing ward-data for the EU Referendum, we define as training data the $280$ ward outcome data available for the three elections and as the test data for the ME's as the $328$ ward outcomes that are missing for the EU Referendum, that correspond to a $53.94\%$ of overall wards (the training data is thus smaller than the test data). Since the EU Referendum outcomes are not available at the ward level for the test wards,  we thus evaluate the prediction on the corresponding $18$ Borough outcomes. We perform the prediction for the best (the lowest distance for the training data) $T=100$ parameter sets, and evaluate the prediction performance in terms of the average (over $100$) distance on the test data 
$\langle \eta(S'_{test},S_{test}) \rangle=\frac{1}{T}\sum_{t}^{T} \frac{\sum_{c \in 
test} n_c \left|{S'}_c^t-S_c \right|}{\sum_{c \in test} n_c}$, where $S_c$ and 
$S'_c$ are the real and the predicted outcomes of $c$ ward/Borough, and $n_c$ is the 
number of spins in $c$. We compare the predictions of our KBI model ($J>0$) with the predictions of a model with only External Fields and without social connectivity kernel ($J=0$, Only EFs) (see results of ABC marginal posteriors of the model parameters with $J=0$ for the three elections in SM Fig. S5-S6). The top $100$ accepted simulations had an error of $\eta(S',S)<0.078$ ($J>0$) vs $\eta(S',S)<0.083$ ($J=0$) in the Mayoral election and $\eta(S',S)<0.062$ ($J>0$) vs $\eta(S',S)<0.067$ ($J=0$) in the EU election: we thus find support favouring the richer KBI model for both types of votes. We also compared the predictions with a baseline Null Model that assigns the same prediction for all outcomes in the test, which is the weighted average outcome on the training data, $S_{pred}=\frac{\sum_{c \in training} n_c S_c}{\sum_{c \in training} n_c}$. 

Results in Fig. \ref{fig:predictability} show that the KBI model outperforms predictions with only EFs in all the three training tests. In particular, we observe larger improvements for Mayoral Elections than for the EU Referendum, where the relative advantage when considering the connectivity kernel is smaller. We believe this is due to differences in the ward-polarisation of the electoral outcomes, where we define polarisation as the tendency of a distribution of ward-level vote-shares to have separated peaks \cite{esteban94}. In SM Fig. S8, we show that Mayoral Elections have larger ward outcome polarisation than the EU referendum. According to our experiments also in Fig. S8, for the same model parameters by removing the connectivity kernel the MEs outcomes distribution dramatically increase their polarisation while for EU Referendum outcomes distribution remain significantly less affected. This suggests that for Brexit the kernel adds less new information above the EFs compared to the MEs, as we would expect with outcomes that do not have clearly defined niches in the Blau space \cite{mcpherson19}. Overall, we demonstrate that the model can successfully predict missing electoral data.




\subsection{Reducing homophilic tendencies and social inequality can reduce polarisation \emph{in silico}}
\label{sec:Optimization}

\begin{figure}[t!]
\includegraphics[width=\columnwidth]{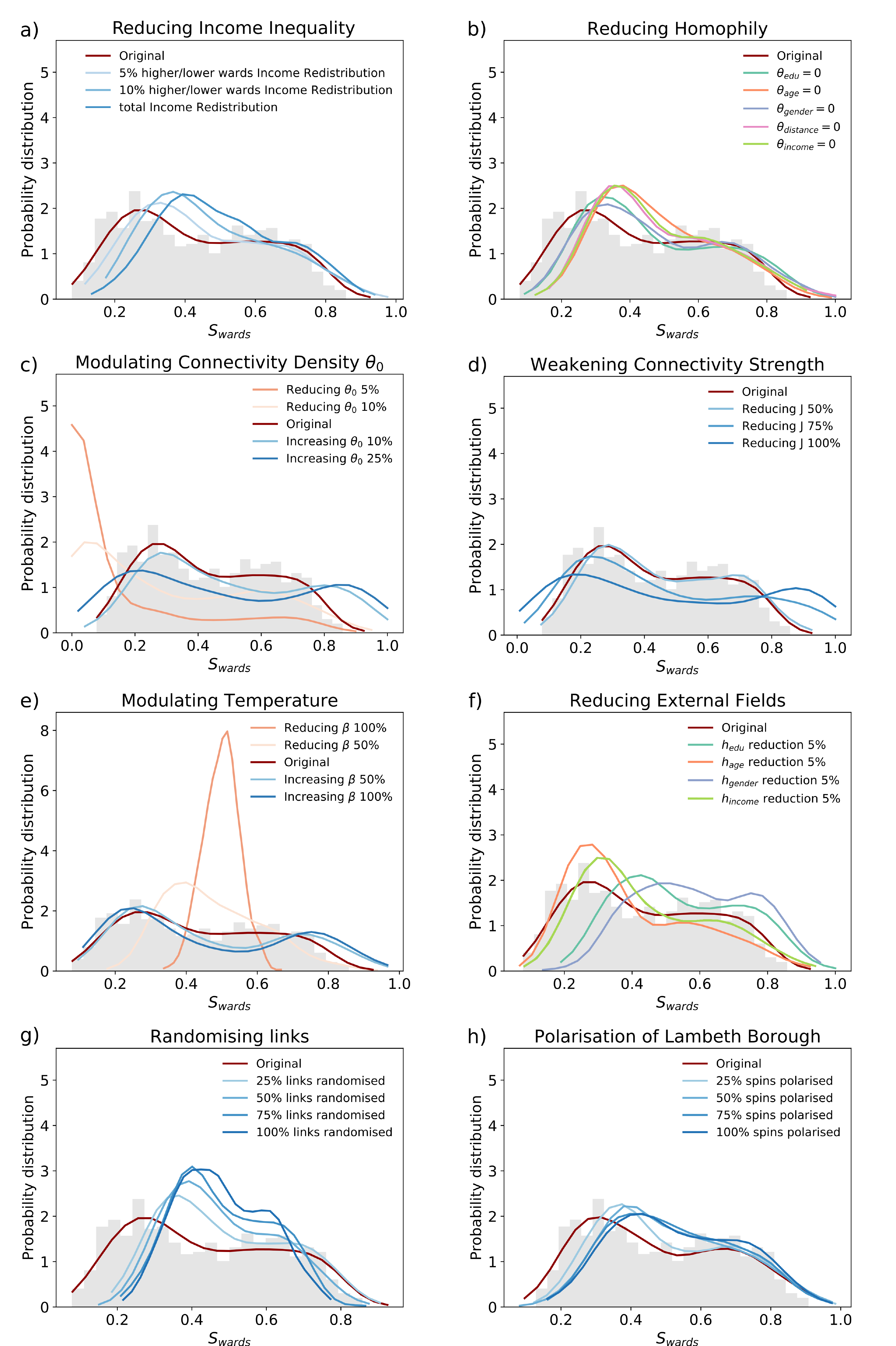}
\caption{\textbf{Model polarisation can be reduced by reducing inequalities and encouraging social mixing.} For simplicity, we only show results for the 2016 London Mayoral Election. On the x-axis, $S_{ward}$ is the fraction of Conservative spins in the wards, being $S_{ward}=0$ all votes for Labour and $S_{ward}=1$ all votes for Conservatives. We show in grey the histogram of the real electoral outcomes in the $608$ electoral wards and in dark-red the Original Gaussian kernel density of electoral outcomes from a multivariate Gaussian of ABC marginal posteriors in Fig. \ref{fig:Bremay280}. In solid lines the different Gaussian kernel density estimates for the different interventions in the system. 
\label{fig:Control}
}
\end{figure}
The interpretability of our model parameters allows us to explore different notional scenarios for public intervention while illuminating the role that the different elements of the model play in producing social outcomes. In this section we illustrate how possible interventions might reduce electoral polarisation. This is, of course, an \emph{in silico} exercise which, dramatically, supposes the microscopic and causal relevance of the model we fit. We define electoral ward outcome polarisation as the tendency of the distribution of ward outcomes $S_{ward}$ (as the fraction of votes to one of the two options) to be peaked towards the extremes $S_{ward}=0$ and $S_{ward}=1$ of the distribution as $\mathscr{P}(S)=\frac{1}{N_L}\sum_{(i,j) \in L} |S_i -S_j|$, where $S_i$ is the outcomes of ward $i$, and $L$ and $N_L$ are all pairs and the number of pairs of electoral wards respectively. In Fig. \ref{fig:Control} we to use the wards outcome distribution only for 2016 London Mayoral Election for simplicity, although the observed effects in Fig. \ref{fig:Control} should be qualitatively similar for 2012 ME and the EU Referendum elections given that they also present homophilous kernels and similar thermal noise and connection strengths. As a starting point, in Fig. \ref{fig:Control} grey histograms showing the observed data, we see that far from a Gaussian distribution the ward outcomes distribution is polarised, with a group of electoral wards voting by majority Labour, then a decrease in the number of wards voting 50/50 and again a higher number of wards voting by majority Conservatives (see Fig. S8 for the probability distributions of $S_{ward}$ for the three elections and their polarisation). In what follows we show that social outcome polarisation can be reduced \emph{in silico} following very different strategies: by reducing inequality (changing the Blau-space coordinates of individuals) (a), by reducing homophily (changing the connectivity kernel)(b,g), by increasing thermal noise (e), by shifting the EFs (f), or by shifting single/multiple wards outcomes (directly altering the configuration of the spins) (h). 

Sub-figure (a) shows how redistribution of income ---changes in the coordinates of wards in the Blau space--- for those wards with lower and higher income reduces polarisation. Selected wards' income are set to their weighted average so that the total income in conserved. Importantly, only redistribution of income of the $30$ wards with the lowest income and the $30$ wards with the highest income in Greater London (corresponding to the $5\%$ lowest and $5\%$ highest income wards) results in a dramatic drop in polarisation. \emph{In silico} results indicate that even small efforts reducing inequality can have a big impact in terms of social outcomes. In sub-figure (b) we eliminate homophily in the different Blau space dimensions: effectively eliminating social preference based on these features. While the external fields alone (without considering network effects) are strong predictors of voting outcomes (Fig. \ref{fig:predictability}) we find that modulating homophily can also significantly shift outcomes in the model. As could be expected, the distributions are less polarised compared to the original, for age, distance and income which are the dimensions with stronger homophilous signal (original distribution polarisation, in dark red, is of $p=0.227$, then removing homophily in the different dimension from less polarised to more polarised: removing homophily in age $p=0.190$, income $p=0.197$, distance $p=0.203$, and then gender and education $p=0.223$). 
In sub-figure (c) 
we also show the changes in the 2016 ME outcomes when changing link density and keeping the homophilous kernel parameters. What we see is that a small decreases of the bias term $\theta_0$ ($5-10\%$ reduction) results in a network where connectivity is high enough that almost all nodes align towards Labour vote. On the other hand, when links' density decreased the mixing effect of the network is reduced resulting in a more polarised scenario. We observe a similar effect in sub-figure (d) for decreasing link-strength. Reducing the network effect results in a decrease of the social mixing of outcomes, giving rise to more polarised outcomes. Therefore, even though the connectivity network is homophilous, and reducing homophily decreases polarisation, the network also serves to mix population voting behaviour and thus weakening the network increases polarisation. 
Modulating thermal noise in sub-figure (e) has a strong effect on the outcomes. Increasing thermal noise (reducing inverse temperature $\beta$) decreases polarisation and the other way around, cooling down the system increases polarisation. As might be anticipated from conventional social science models, we observe big changes when manipulating the linear External Fields sub-fig (f). The changes in the external fields shifted the outcome distribution towards more negative or more positive outcomes, while reducing polarisation. In sub-figure (g) we reduce homophily by randomising links. Clearly, the randomisation of an homophilious network increases the mixing of outcomes and reduces polarisation. Finally, we can also change polarisation by shifting opinions locally. In sub-figure (h) 
we fix the alignment of increasing fractions of spins in the Borough of Lambeth, a traditionally Labour Borough ($72\%$ vote to Labour) towards Conservative vote, and measure the effect on the other wards outcomes (Lambeth Borough has 21 wards, corresponding to a $\sim4\%$ of all spins in the system). For the outcomes distribution in this sub-figure we remove outcomes from Lambeth Borough wards, thus we only measure the changes on the remaining wards. The changes in the ward outcomes distributions are only due to the fact that wards are connected. We see that polarisation decreases and the outcomes are shifted towards more Conservative outcomes. 

\section{Discussion} 
We presented the kernel-Blau-Ising model that allows us to accurately reproduce population level behaviour and known homophilic tendencies only using snapshot population behavioural data, without using network data. By effectively inferring the metric for behaviourally relevant social connections in the Blau space we are not only able to improve our understanding of the behavioural process and its links to inequalities and social preferences, but also to improve predictability of unobserved social outcomes.

Our KBI model exploits the spatial character of homophilic preferences to reduce the number of parameters needed, thus allowing us to perform the inference from snapshot data and to avoid needing time-series data as in \cite{peixoto19,schaub19,fedele13b}. Our efforts have much in common with other recent work that seek to merge the notion of behaviour in Blau space with social networks through social network survey data \cite{hoffmannthesis,mcpherson19}. In our case we avoid using any social network survey data, instead recovering a network model tuned to each behaviour type. The model is designed to avoid confounding of homophily and behaviour \cite{shalizi11} but does so by making strong model assumptions about the form of the EFs and their interplay with social connections. Investigating its appropriate generalisations is for further work.


Our illustrative results on voting data in Greater London corroborate known homophilic tendencies for social ties as regards distance, age, income and education \cite{McPherson01,smith14,wang13}. Spatial distance is consistently the most homophilic dimension for the formation of social ties for the three votes, which is in agreement with previous studies \cite{hipp09,hoffmannthesis} and, further, the spatial length scales we infer are consonant with spatial homophily
observed by others \cite{hoffmannthesis,mcpherson19}. Interestingly, we found large similarities between the inferred social network structures for 2012 and 2016 London Mayoral Elections while a different social structure for the EU (Brexit) Referendum. This is consistent with the observation that traditional left--right politics do not help explain the Brexit Referendum vote \cite{swales16,johnston18,hobolt18} perhaps suggesting that people discussed Brexit with a different type of person from the ones with whom they discussed the MEs \cite{cowan17}.  Regarding the apparent differences between votes, the EU Referendum fit (unlike the Mayoral elections) shows educational homophily: the models that were selected suppressed inter-educational communication. Moreover, we do not infer the significant age or income homophily that we observe for both Mayoral Elections. Education has been particularly highlighted as a significant explanatory variable (though treated as an external field) in other studies \cite{becker17,alabrese19}. 

Because the model and its parameters are interpretable, it enables us to explore, \emph{in silico} different intervention strategies in a novel framework. Beyond observing that an \emph{in silico} reduction in income inequality reduces polarisation, we also find unexpected strategies: in the model of the 2016 Mayoral Election we achieved a larger depolarising effect by allowing more inter-generational connectivity --eliminating age homophily-- than by eliminating income or distance homophily. This hypothesis is suggestive since reducing age homophily might be less contentious than other social interventions. Interestingly, we found that although the inferred social network is homophilous, it nonetheless helps to mix population outcomes: weakening the social network effect --by decreasing connectivity density or connection strength-- results in a more polarised scenario. 

Given its simplicity and interpretability, we believe that KBI model will be useful in a wide variety of behavioural/attitudinal processes where the social network appears to have an effect, with special focus on health risk behaviours such as smoking, alcohol consumption or vaccine refusal. We also believe the KBI model is relevant since it makes explicit, in a simple and interpretable manner, how social inequalities (from income to educational inequities) interplay with our social preferences to shape social network structure; and then how in turn, our social networks and social tendencies (external fields) shape our behaviours. 

\begin{acknowledgments}
This work was supported by the EPSRC Centre for Mathematics of Precision Healthcare (EP/N014529/1).
\end{acknowledgments}


\setcounter{equation}{0}
\renewcommand\theequation{A.\arabic{equation}}


\bibliography{ref_mine}

\end{document}


\renewcommand{\thefigure}{S\arabic{figure}}
\renewcommand{\theequation}{S\arabic{equation}}
\renewcommand{\thetable}{S\arabic{table}}
\setcounter{figure}{0}
\setcounter{equation}{0}
\setcounter{page}{1}


\begin{center}
	\huge
	Inference and Influence of Large-Scale Social Networks Using Snapshot Population Behaviour without Network Data\\
	\medskip
	\Large
	(Supplemental Material)\\
	\medskip
	Antonia Godoy-Lorite, Nick S. Jones
\end{center}


	%

\section{Datasets}
\label{sec:Datasets}

The electoral outcomes we use in Section IV in the main text are London Mayoral Election 2012 and 2016 from http://data.london.gov.uk/elections, and the EU Referendum for Greater London from the BBC's electoral wards outcomes at https://3859gp38qzh51h504x6gvv0o-wpengine.net\\dna-ssl.com/files/2017/02/ward-results.xlsx. For Mayoral Elections we only use Conservative and Labour votes to keep to binary outcomes, since they account for the $84\%$ of 2012 ME and $79\%$ 2016 ME. We use data at the electoral ward level, given that it is the smallest spatial resolution that we get for electoral outcomes, such that Greater London is divided into $630$ electoral wards with an average electorate of $7750\pm1550$ each. The EU Referendum ward level data are incomplete, there are $18$ boroughs (out of 32 boroughs excluding the City of London) that did not release the outcomes at the electoral wards level but aggregated them at Borough level, which are: Bexley, Brent, Barnet, Barking and Dagenham, Newham, Tower and Hamlets, Sutton, Hillingdon, Lewisham, Redbridge, Southwark, Wandsworth, Hackney, Kingston, Richmond, Hammersmith and Fulham, Kensington and Chelsea, and Westminster. For those Boroughs we are using the aggregated outcomes at Borough level at http://data.london.gov.uk/electi\\ons, also for 2012 and the 2016 London Mayoral Elections in Figs.\ref{sf-EF280}-\ref{fig:may608}. 
Regarding the Blau space coordinates of each ward, we compute the average for each of the chosen dimension of the Blau space, which are: education (categorical variable with 5 categories), age (ordinal variable in binning of 1 year from 18 to 100 years), gender (binary variable 0 for male and 1 for female) all of them from census data 2011 https://www.ons.gov.uk/census/2011census, http://infusecp.mimas.ac.uk/, then spatial location (centroid coordinates of each ward) and income in 1000GBP (median household income estimates from 2012/2013 https://data.london.gov\\.uk/dataset/ward-profiles-and-atlas). The Blau space coordinates are not standardised --only the relative distances are standardised-- but we use age and income coordinates multiplied by a factor $0.1$ and gender by a factor $10$ so that they are of the same order of magnitude. The City of London has an special treatment inside Greater London, it is not considered a Borough as the other 32 boroughs in GL and there is no census data on it, thus we did not include it in our analysis. Also due to a miss-match between the definition of electoral wards used for electoral outcomes and 2011 census wards, we could only use $608$ wards out of the $630$ electoral wards and the $625$ 2011 census wards.

\section{Inferring model parameters in Synthetic data}
\subsection{Rescaling population size}

For computational expedience our treatment of electoral data uses a subsample of the full population. Here we make the case for approximating systems of size $N^{(1)}$  with smaller systems of size $N^{(2)}$. We suggest that inference of the kernel parameters (except the bias parameter $\theta_0$) can be insensitive to changes in the population size $N$. We demand that the update dynamics of spin $i$ be conserved under system-size rescaling where, each Blau co-ordinate $z$, has spin-up fraction $S_{z}$. The External Field's effect on each spin’s update dynamics does not change when the system size is rescaled and so can be neglected. Single-spin update dynamics rescaling thus depends only on the term $\sum_j A_{ij}\sigma_j$: it suffices to show that this term is unchanged under a system size rescaling (when combined with a rescaling of $\theta_0$).

If the number of spins at each Blau-space co-ordinate is sufficiently large then $\sum_j A_{ij}\sigma_j \sim \sum_z \rho_{iz}S_z n_z$ where $n_z$ is the number of spins at co-ordinate $z$ and $\rho_{iz}$ is the chance of connecting $i$ to any node at location $z$. If we require that when we approximate systems of size $N^{(1)}$  with smaller systems of size $N^{(2)}$ we have: a) $n_z^{(1)}$ and $n_z^{(2)}$ are related by $\frac{n_z^{(1)}}{n_z^{(2)}}=\frac{N^{(1)}}{N^{(2)}}$ (i.e. we rescale the population size of each Blau-space co-ordinate in a manner proportional to the overall system-size) and b) that $\theta_0^{(2)}=\theta_0^{(1)}-log(N^{(1)}/N^{(2)}) ,\: \theta_k^{(2)}=\theta_k^{(1)} \: k\neq 0$, then we find that $\sum_z \rho^{(1)}_{iz}S_z n^{(1)}_z = \sum_z \rho^{(2)}_{iz}S_z n^{(2)}_z$. This result follows straightforwardly given requirement a) and because requirement b) ensures $\rho^{(2)}_{iz}=\frac{N^{(1)}}{N^{(2)}} \rho^{(1)}_{iz}$. Given that $\sum_z \rho^{(1)}_{iz}S_z n^{(1)}_z = \sum_z \rho^{(2)}_{iz}S_z n^{(2)}_z$ the update dynamics of each spin are unchanged under rescaling.

We now justify why $\theta_0^{(2)}=\theta_0^{(1)}-log(N^{(1)}/N^{(2)})$ implies $\rho^{(2)}_{iz}=\frac{N^{(1)}}{N^{(2)}} \rho^{(1)}_{iz}$. $A_{ij}$ is Bernoulli distributed according to the kernel $\rho(i,j)=\frac{1}{1+exp(\theta_0 + \sum_k \theta_k|z_{ik}-z_{jk}|)}$.  For $N^{(1)}\gg1$ and $N^{(2)}\gg1$, but small finite expected degree, the connection probabilities are low so that $\rho(i,j)=\frac{1}{1+exp(\theta_0 + \sum_k \theta_k|z_{ik}-z_{jk}|)} \: \approx \:  e^{-\theta_0}e^{-\sum_k \theta_k |z_{ik}-z_{jk}|}$, therefore if $\theta_0^{(2)}=\theta_0^{(1)}-log(N^{(1)}/N^{(2)})$, then  $\rho^{(2)}\approx \frac{N^{(1)}}{N^{(2)}} e^{-\theta_0^{(1)}}e^{-\sum_k \theta_k^{(1)} |z_{ik}-z_{jk}|}\approx\frac{N^{(1)}}{N^{(2)}}\rho^{(1)}$. 

As seen in Fig. S1 the approximation yields similar inference results, as regards the maxima (and general form) of the posteriors. A consequence of using a smaller system to represent a larger is that the smaller system can show proportionately larger fluctuations; this can inflate the variance of the posteriors leading to an overestimate in parametric uncertainty. A comparison of the  right and left sides of Fig. S1 shows that the smaller system has slightly wider posteriors. While suitable for our illustrative objective here, this is a direction that requires further optimisation should sharper posteriors be required.

\begin{figure}[h!]
	\centerline{
		\includegraphics*[width=\columnwidth]{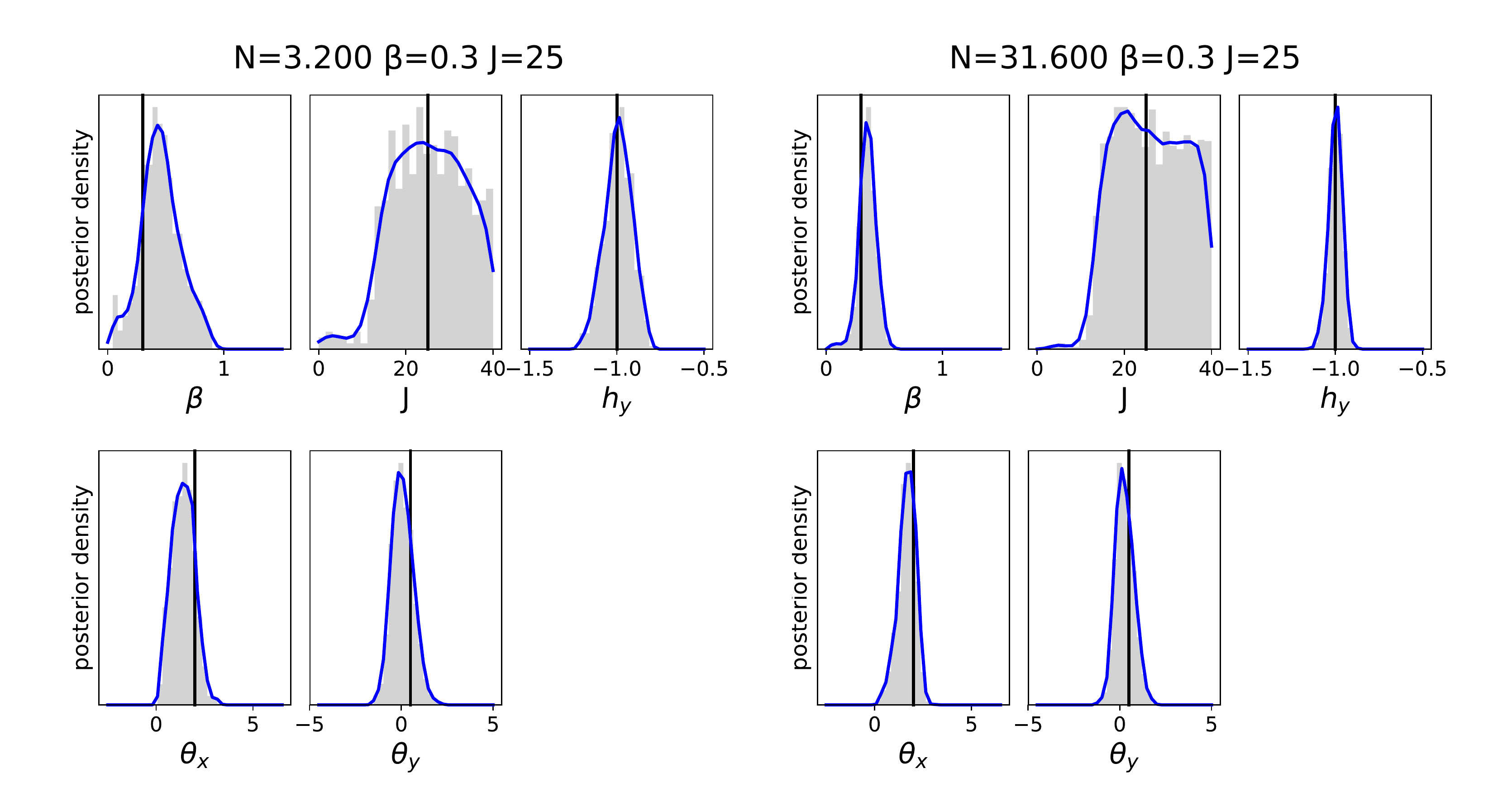}
	}
	\caption{{\bf Dimension-specific kernel parameters are insensitive to a rescaling of the system size}. As input data for the inference we use a single-observation of the summary outcomes $S_z$ for $z=(x,y) \in x,y=\{0,1,..,9\}$ generated for $N=10,000$ for Fig. 2 in the main text with $\theta_0=9$ (only for $J=25$). We show the ABC marginal posteriors (for $500$ samples with the lowest distance $\eta(S',S)$) for two different system sizes: on the left figure we show the ABC marginal posteriors for a smaller system with $N=3,200$ ($\theta_0=7.86$) and on the right for a larger system with $N=31,600$ spins ($\theta_0=10.15$). We show that given the same single-observable outcomes we can re-scale the system size and recover indicative model parameters, and in particular the dimension-specific kernel parameters which are insensitive to the re-scaling.
	}
	\label{sf-N}
\end{figure}

\newpage

\subsection{Degeneracy between network degree and the connection strength}
To illustrate the degeneracy between the bias term $\theta_0$ and the connection strength in the Hamiltonian (Eq.3 in the main text), we perform the inference of the model parameters for an observable generated with $\theta_0=9$ ($\langle \kappa \rangle \approx 2$) but now sampling spin configurations with $\theta_0=7$ ($\langle \kappa \rangle \approx 10$) for a much higher density of connections. 


\begin{figure}[h!]
	\centerline{
		\includegraphics*[width=\columnwidth]{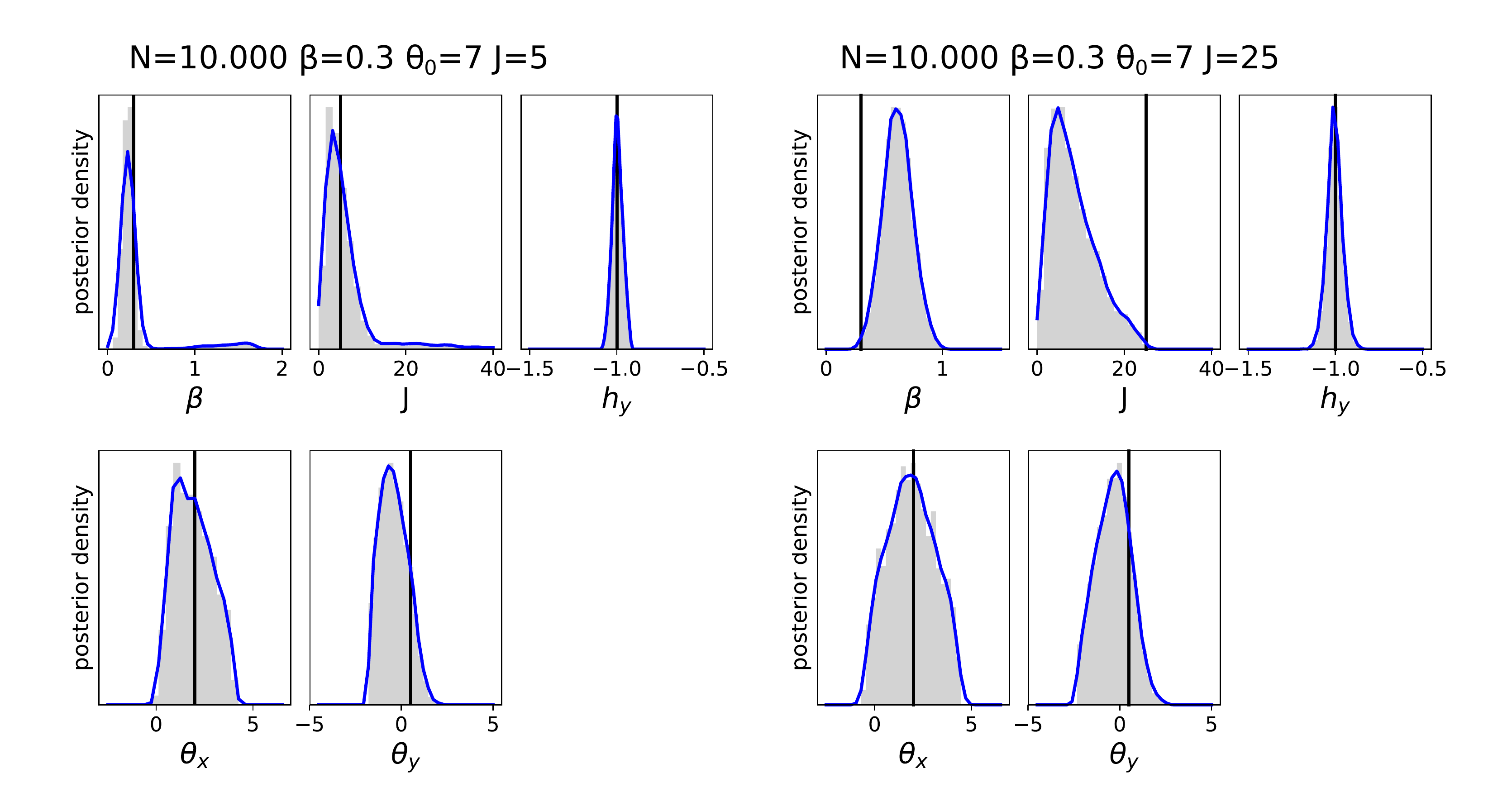}
	}
	\caption{{\bf Dimension-specific kernel parameters are insensitive to changes in the degree of the network}.  As in Fig. S1, we use as input data for the inference the single-observation of the summary outcomes $S_z$ for $z=(x,y) \in x,y=\{0,1,..,9\}$ generated for Fig. 2 in the main text (with $N=10,000$ and $\theta_0=9$) for weak connection strength ($J=5$) and strong connection strength ($J=25$). We show the ABC marginal posteriors for $500$ samples with the lowest distance $\eta(S',S)$ for both scenarios instead assuming a higher connectivity density $\theta_0=7$, corresponding to an average degree of $\langle \kappa \rangle \approx 10$. We found that the connection strength decreases when increasing the density of connections. However, although ABC marginal posteriors result in wider distributions, we are able to extract properties of the dimension-specific kernel parameters, $\theta_x, \theta_y$, from systems with higher connectivity density. 
	}
	\label{sf-th0}
\end{figure}
\newpage


	%
\section{Synthetic and electoral data magnetisation}

\begin{figure}[h!]
	\centerline{
		\includegraphics*[width=\columnwidth]{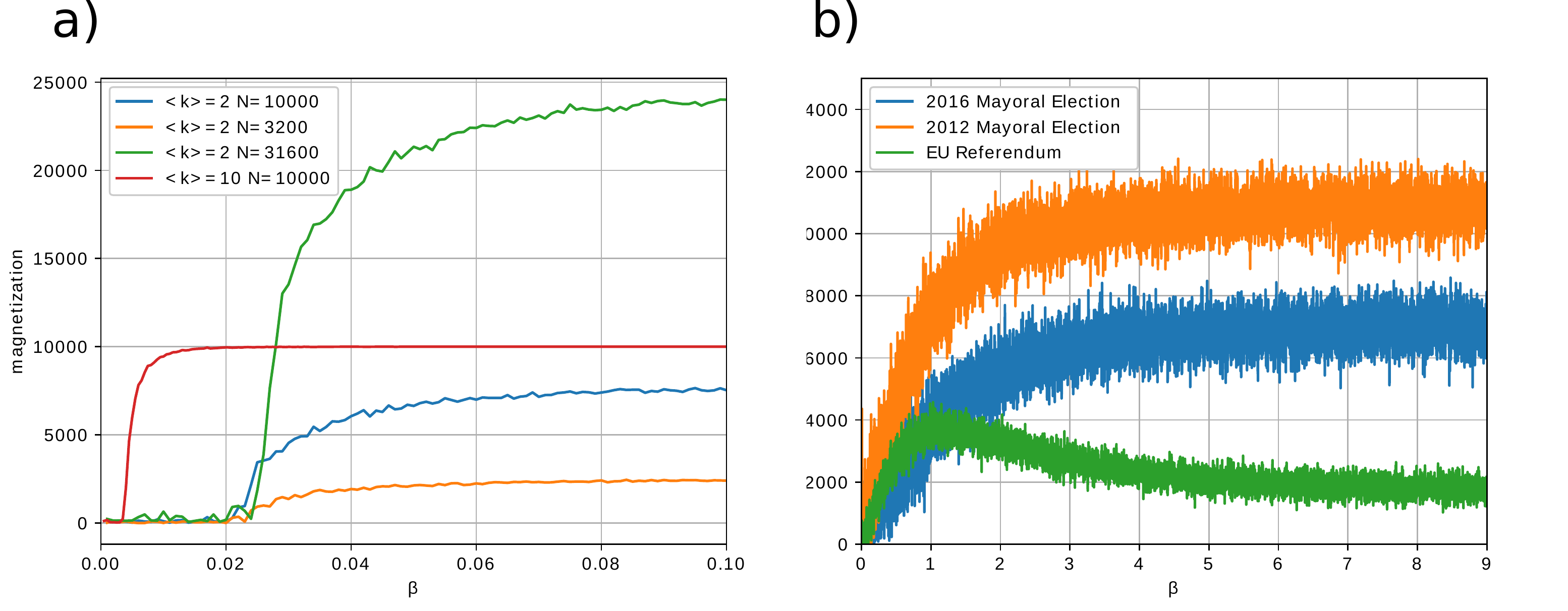}
	}
	\caption{{\bf Magnetisation both for synthetic data and for Greater London electoral data}. {\bf a} We show the spins magnetisation on synthetic data for different system sizes and different connectivity density kernels (see the legend), but same EFs ($h_x=1,:\ h_y=-1$). {\bf b} Spins magnetisation for 2012 and 2016 London Mayoral Elections and EU Referendum. Except for $\beta$, the model parameters used to generate the different spin configurations are from a multivariate Gaussian fit on the ABC marginal posteriors in Fig. 3 in the main text.
	}
	\label{sf-syntheticmagnetization}
\end{figure}

	%
\newpage
\section{Bootstrapping population Blau space coordinates from Census micro-data}

\begin{figure}[h!]
	\centerline{
		\includegraphics*[width=\columnwidth]{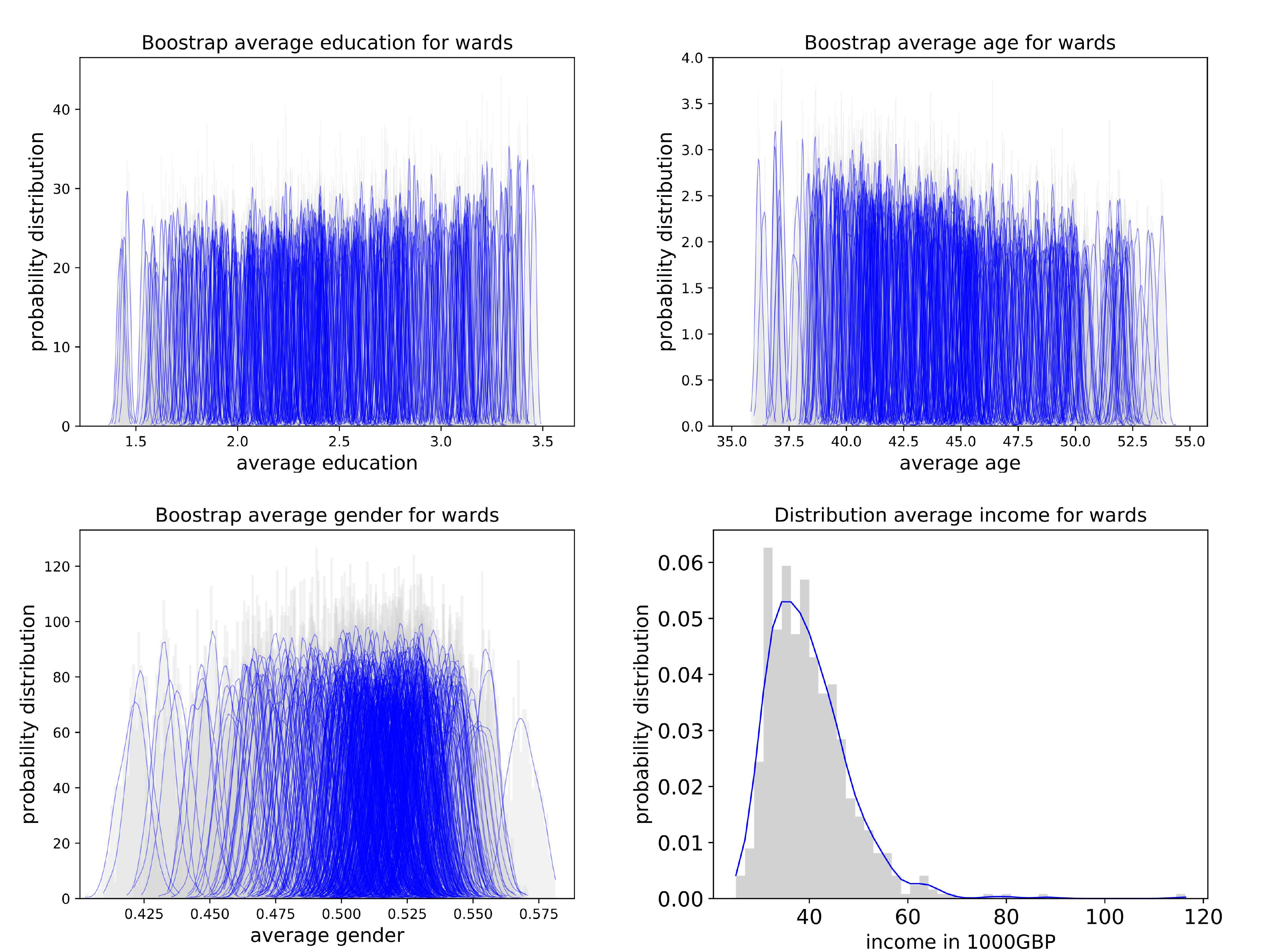}
	}
	\caption{{\bf  Evidence of heterogeneous distribution of average wards’ coordinates by bootstrapping census micro-data at the ward level}. The different sub-figures show the bootstrapping averages for the different $608$ wards in Greater London from 2011 census data distribution on education, age and gender for $1,000$ samples of the actual ward population size ($\sim 7750$ individuals). 2011 census data does not include an income variable, hence, we do not have ward level micro-data but only ward level income estimates (see Section VI in the main text), therefore, we show the overall probability distribution of income for the $608$ wards in GL. 
	%
	}
	\label{sf-boostrapping}
\end{figure}
\newpage
\section{Multiple Linear Regression on electoral data}
\label{sec:MLR}
In order to get some simple estimates of what could be expected from a simple linear regression on the Blau space, we perform a multiple linear regression for 2012 and 2016 MEs and the EU Referendum electoral outcomes with education, age, gender and income as independent variables (we exclude geospatial $(x,y)$ Blau coordinates from the linear regression). The Blau space coordinates are not standardised -- only the distances are standardised-- but we use age and income coordinates multiplied by a factor $10$ and gender by a factor $0.1$ so that they are of the same order of magnitude:

\begin{table}[h!]
\begin{tabular}{lllll}
 & education & age & gender & income \\
 2012 ME &  0.460 & -0.333 & 0.489 & -0.553 \\
 2016 ME & 0.459 & -0.491 & 0.583 & -0.457 \\
 EU Referendum & 0.453 & -0.376 & 0.177 & -0.033
\end{tabular}
\end{table}

\section*{}
\section{ABC inference only with External Fields ($J=0$)}
%
%
We show the ABC marginal posteriors for only External Fields (without connectivity kernel) for different input data: for 2012 and 2016 London Mayoral Elections with $608$ electoral wards outcomes; and given the missing wards' data for EU Referendum (see Section VI in main text), we use the available $280$ ward level data and we use Borough level electoral outcomes for the remaining $18$ Boroughs, for EU Referendum, and in order to compare the estimated model parameters with exactly the same input data, also for the 2012 and 2016 MEs.
\begin{figure}[h!]
	\centerline{
		\includegraphics*[width=0.8\columnwidth]{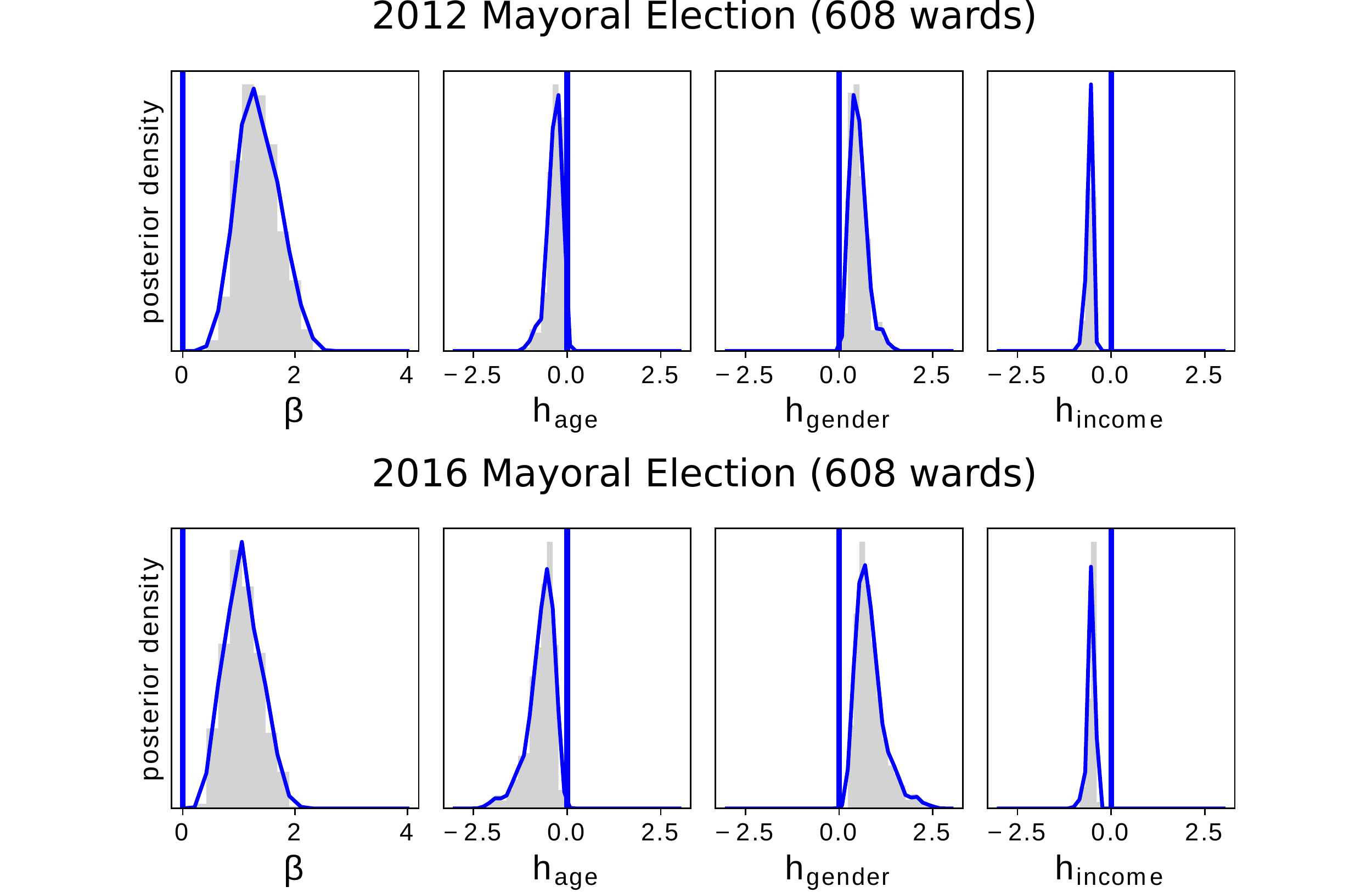}
	}
	\caption{{\bf Only External Fields inference for 2012 and 2016 London Mayoral Elections with 608 electoral wards}. We show the ABC marginal posterior distribution for the model parameters excluding the connection term (only using linear External Fields) for $500$ samples with the lowest distance $\eta(S',S)$-- such that all samples have a distance (or WMAE) $\eta(S',S)<0.098$ for both Mayoral Elections. Given the degree of freedom in the Hamiltonian (Eq.3 in the main text), we choose to fix the education coefficient to $h_{edu}=0.45$. We find that although 2012  ME External Fields are peaked slightly closer to zero than 2016 ME External Fields estimates, overall the estimates are similar for the two Mayoral Elections. 
	}
	\label{sf-EF608}
\end{figure}
\begin{figure}[h!]
	\centerline{
		\includegraphics*[width=0.8\columnwidth]{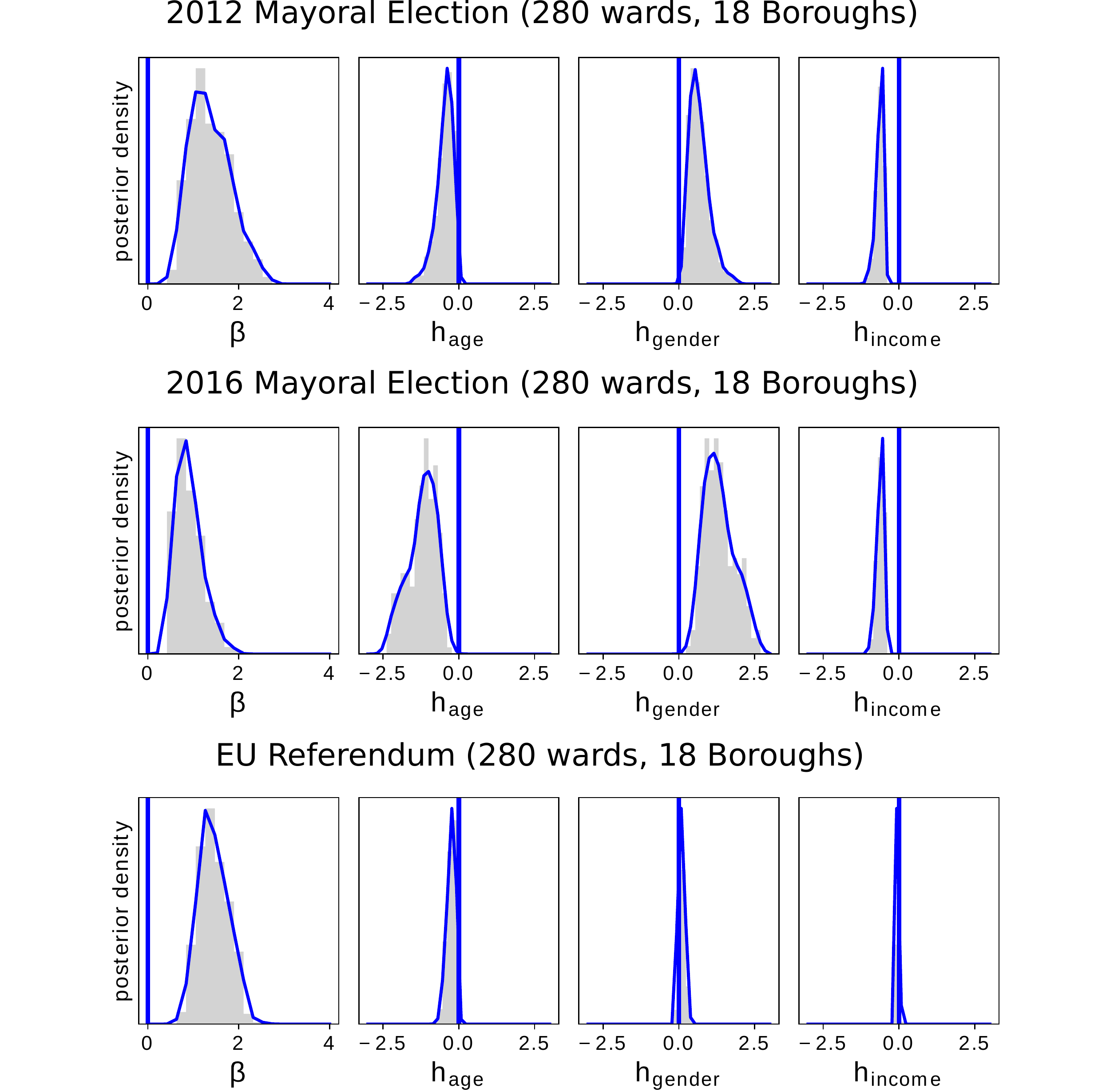}
	}
	\caption{{\bf Only External Fields inference for 2012 and 2016 London Mayorial Elections and EU Referendum inference with 280 electoral wards and 18 Boroughs}. We show the ABC marginal posterior distribution for the model parameters excluding the connection term (only using linear External Fields) for $500$ samples with the lowest distance $\eta(S',S)$-- such that all samples have a distance (or WMAE) $\eta(S',S)<0.083$ for both Mayoral Elections and $\eta(S',S)<0.052$ for the EU Referendum. Given the degree of freedom in the Hamiltonian (Eq.3 in the main text), we choose to fix the education coefficient to $h_{edu}=0.45$. We find EFs estimates are  more peaked around zero for the EU Referendum. 
	}
	\label{sf-EF280}
\end{figure}

\newpage


\section{ABC inference on electoral data with 280 wards and 18 Boroughs as input data}
\begin{figure}[h]
\includegraphics[width=\columnwidth]{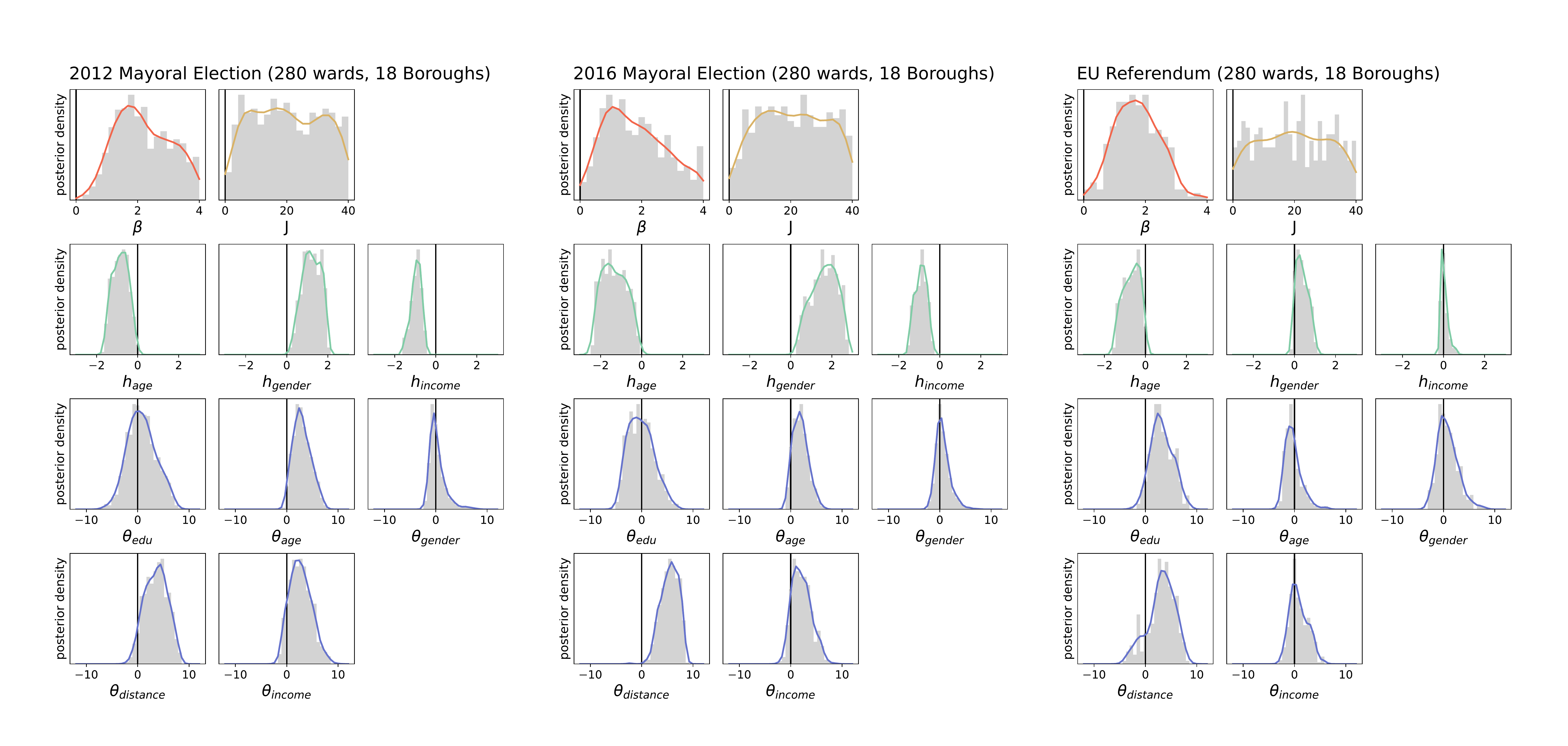}
\caption{\textbf{ABC marginal posteriors for 2012 and 2016 London Mayoral Elections and  the EU referendum using 280 electoral wards and 18 Boroughs as input data}. As a complementary figure to Fig. 3 in the main text, here we show the ABC parameters estimates for the three elections with the same input data as EU Referendum data (280 ward outcomes and 18 Borough outcomes). We show the ABC marginal posteriors for $500$ samples with the lowest distance $\eta(S',S)$ -- such that all samples have a distance (or WMAE) $\eta(S',S)<0.071$ for both Mayoral Elections and $\eta(S',S)<0.050$ for the EU Referendum. In the inference procedure, we choose to fix $\theta_0=14$ and $h_{edu}=0.45$ given the degeneracies of the model (Eq.4 in the main text). We show the histograms of the ABC marginal posteriors in grey, as a solid line the Gaussian kernel density estimates, and the vertical lines correspond to $0$ value. The parameters estimates are almost the same for these input data as for using $608$ wards input data in Fig. 3 in the main text, with distance being the strongest homophilous signal and age and income also showing significant homiphilic signal. 
\label{fig:may608}
}
\end{figure}
\newpage
\section{Electoral outcomes distributions and polarisation}

In this section we show the polarisation of the different electoral ward outcomes distributions for the KBI generated outcomes with and without connections (connectivity kernel). We measure polarisation of wards outcomes as in J.M Esteban and D. Ray (2007)-- without considering group identification-- such as,
\begin{equation*}
    \mathscr{P}(S_w)=\frac{1}{N_L} \sum_{(i,j) \in L} |S_i -S_j|,
\end{equation*}

where $S_i$ is the electoral outcome of ward $i$ given as the fraction of votes to one of the two options, $L$ and $N_L$ are all pairs of electoral wards (for $i\neq$j) and the number of pairs of wards respectively. Notice the difference between measuring polarisation between wards' outcomes and polarisation inside wards or of the overall vote. Wards' outcomes polarisation would be zero if all wards have the same outcomes regardless if that is $S_w=0, \: S_w=1$ or $S_w=0.5$, while overall votes polarisation would be $0$ for $S_w=0, \: S_w=1$ and maximum for $S_w=0.5$. We are focusing on the differences between ward-level outcomes as a proxy for the degree to which wards have similar voting composition to others, but not how far from consensus is the population vote. In what follows, we compare polarisation between the different electoral ward outcomes distributions from the ABC estimates in Fig. 3 in the main text, and for the generated outcomes with the same parameters but eliminating the connectivity kernel (with J=0).
\begin{table}[h]
\caption{Electoral outcomes Polarisation}
\begin{tabular}{llll}
 2012 ME &  0.239 & 2012 ME with J=0 & 0.384 \\
 2016 ME & 0.228 & 2016 ME with J=0 & 0.351  \\
 EU Referendum & 0.163 & EU Referendum with J=0 & 0.186 
\end{tabular}
\end{table}
\begin{figure}[h]
	\centerline{
		\includegraphics*[width=\columnwidth]{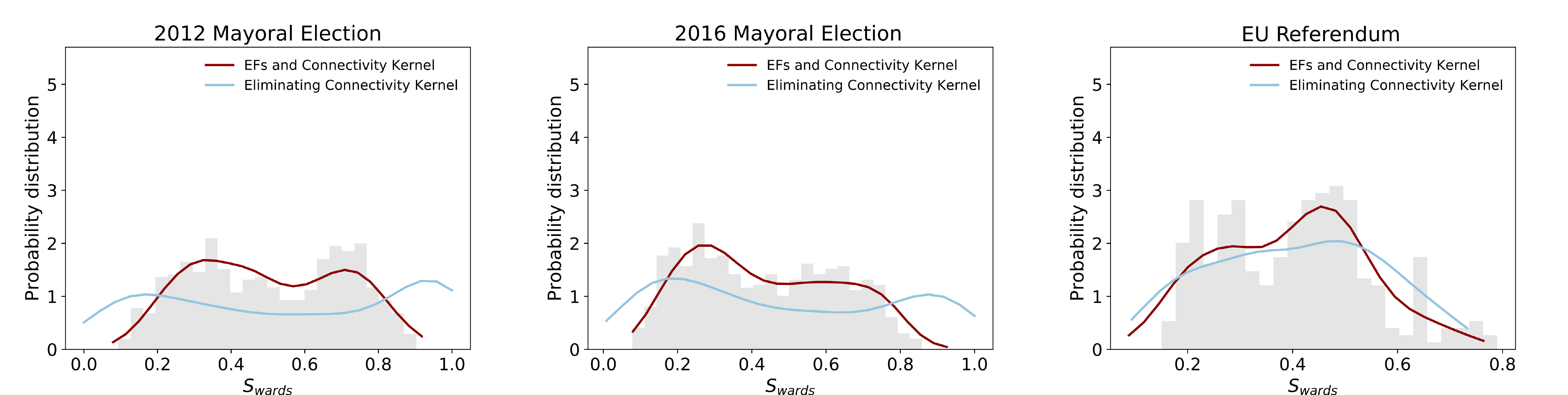}
	}
	\caption{{\bf Electoral outcomes distributions and polarisation}. In the x-axis $S_{ward}$ is the fraction of Conservative/Leave votes in electoral wards, such that  for $S_w=0$ all votes are for Labour/Remain and for $S_w=1$ all votes are for Conservative/Leave. We show in grey the histograms of the real electoral outcomes; in dark-red solid lines the Gaussian kernel density of the KBI generated outcomes distributions from a multivariate Gaussian regression of the ABC estimates in Fig. 3 in the main text; and in light-blue solid lines the Gaussian kernel density of generated outcomes for the same model parameters but eliminating the connectivity kernels (J=0). We find that for the \textit{eliminating connectivity kernel} ($J=0$) distributions, the EU Referendum distribution is in better agreement with the original ward outcomes distribution than the corresponding 2012 and 2016 Mayoral Election distributions. This all suggest that for the EU Referendum the kernel adds less new information above the EFs compared to the MEs, as we would expect with outcomes that do not have clearly defined niches in the Blau space.
	}
	\label{sf-elctionsdistribution}
\end{figure}
\newpage
\bibliography{ref_mine}